\documentclass[fleqn,usenatbib]{mnras}

\usepackage{newtxtext,newtxmath, soul}

\usepackage[T1]{fontenc}

\DeclareRobustCommand{\VAN}[3]{#2}
\let\VANthebibliography\thebibliography
\def\thebibliography{\DeclareRobustCommand{\VAN}[3]{##3}\VANthebibliography}

\usepackage{graphicx}	
\usepackage{amsmath}	
\usepackage{enumitem}
\usepackage{url}
\usepackage{xtab,booktabs, makecell, array}
\usepackage{caption}

\usepackage[normalem]{ulem}

\usepackage[dvipsnames]{xcolor} 

\title[Irbene monitoring I]{Five years of 6.7 GHz methanol maser monitoring with Irbene radio telescopes}

\author[A. Aberfelds et al.]{A. Aberfelds,$^{1}$\href{https://orcid.org/0000-0002-6727-2858}{\includegraphics[scale=0.5]{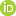}}\thanks{E-mail: artis.aberfelds@venta.lv}
J. Šteinbergs,$^{1}$\href{https://orcid.org/0000-0001-9265-3655}{\includegraphics[scale=0.5]{orcid.png}}
I. Shmeld$^{1}$\href{https://orcid.org/0000-0002-9671-5267}{\includegraphics[scale=0.5]{orcid.png}}
and R. A. Burns$^{1,2}$\href{https://orcid.org/0000-0003-3302-1935}{\includegraphics[scale=0.5]{orcid.png}}
\\
$^{1}$Engineering Research Institute "Ventspils International Radio Astronomy Center", Ventspils University of Applied Sciences, Inzenieru Str. 101, Ventspils, LV-3601, Latvia\\
$^{2}$RIKEN Cluster for Pioneering Research, 2-1 Hirosawa, Wako-shi, Saitama, 351-0198, Japan\\
}

\date{Accepted XXX. Received YYY; in original form ZZZ}

\pubyear{2023}

\begin{document}
\label{firstpage}
\pagerange{\pageref{firstpage}--\pageref{lastpage}}
\maketitle

\begin{abstract}
We present the results of a 5-year monitoring program of 42 sources targeted at 6.7 GHz methanol masers, conducted from March 2017 to October 2022 using the Irbene 32 and 16 meter radio telescopes. Sources were observed with irregular intervals where time between two consecutive observation ranged from twenty four hours to thirty five days. We found that more than 55 per cent of the sources showed significant variability, but often only one or a few spectral features were varying significantly. Numerous type of variability were found in our sample: low-variable, periodic, irregular, synchronised and anti-correlated between features and steadily raising or falling flux. Our analysis techniques also uncover new variability trends for several sources. The maser monitoring program is one of the first single-dish science initiatives at the Irbene radio telescope complex, initiated shortly after the instrument's reconstruction and upgrades. Our findings unequivocally demonstrate its suitability for maser research purposes. 
\end{abstract}




\section{Introduction}

The formation of high-mass stars is still a challenging topic in modern astrophysics with a key question being, how a massive star attains its ultimate mass. Two prevailing theories that tackle this issue are `global collapse' and `competitive accretion' (e.g. \citealt{Zinnecker}, for review).
Studying high-mass star-forming regions (HMSFRs) through observations poses significant challenges due to their typically large distances, obscuration caused by interstellar dust, and rapid evolution. Recently, 6.7\,GHz methanol masers have emerged as a unique and valuable tool for studying HMSFRs, as they serve as prominent signposts of the radiative conditions these regions (e.g., \citealt{menten1991}). The spectral features of these masers are remarkably bright and narrow.
Methanol maser emission has been suggested to appear during the early stages of protostellar evolution \citep{Walsh}.  Most of the observed 6.7 GHz methanol masers are associated with protostellar discs \citep{Walt}, although several cases where the masers trace outflows has also been reported \citep{Goddi_2011}. 

Early insights into 6.7 GHz methanol maser variability were gleaned from surveys focused on discovering new masers. A notable study by \citet{caswell1995} involved observing a sample of 245 sources 4 to 5 times over a 1.5-year period. The results indicated that only 48 sources from the 245 sample showed significant variability, and often only one or several spectral components in these variable sources exhibited high variability. Two epoch (separated by 2 years) multi-beam survey found that 5 to 7 per cent of a 972 source sample showed flux density changes by a factor of two or more \citep{multibeam_catalog}.

The two largest long term methanol maser monitoring surveys according to our knowledge ware conducted by \citet{Goedhart} and \citet{Szymczak}. \citet{Goedhart} reported 6.7 GHz maser flux density changes of 54 sources over a duration of 4.2 years, monitoring with 2 -- 4 observations per month, transitioning to daily observation for rapidly variable sources. \citet{Szymczak} conducted  a 3.7--year monitoring campaign of 166 sources with 2 -- 4 observations per week. 
\cite{Goedhart} classified spectral features as variable if the derived variability index (see section 2.4, equation (1)) was higher than 0.5. \cite{Szymczak} used a different approach based on the variability properties on individual spectral features (see section  2.4 of their paper).   
Nevertheless, \cite{Szymczak} found that fractions of low-variable (and non-variable) sources in both samples are very similar -- 19 and 21 percent accordingly. Furthermore they (\citealt{Goedhart} and \citealt{Szymczak}) concluded that there are two different kinds of the 6.7~GHz methanol maser line variability: flaring (which refers to fastest flux increases in a short time, e.g. flux increase by factor of two in two weeks) and gradual (including periodic) changes of flux densities.

We would like to highlight the Ibaraki 6.7 GHz Methanol Maser Monitoring (iMet) program by \citet{Yonekura2016}. The Japanese team conducts regular monitoring of 442 sources, making it the largest sample for a monitoring campaign by far. Furthermore, they consistently publish their 6.7 GHz methanol maser flux measurements in an open-access database\footnote{\url{http://vlbi.sci.ibaraki.ac.jp/iMet/}}. To our knowledge, only a summary of periodic methanol maser variability has been published \citep{Sugiyama2019}. 

The discovery of periodic variability came as an unexpected finding, with known periods ranging from 24 to 600 days, observed in some or all features. Furthermore, researchers have noticed phase-lags between individual maser lines (\citealt{Goedhart}; \citealt{goedhart2014b}; \citealt{fujisawa2014b} and \citealt{Szymczak_2016}).
The majority of theoretical models addressing periodic masers involve a binary nature of the central object, where the observed changes in maser flux density are modulated by periodic variations of pumping radiation, in response to orbital movements (\citealt{Araya_2010}; \citealt{van_der_Walt_2011}; and \citealt{Parfenov_2014}). However, as one exception, \citet{Inayoshi_2013} proposed a model suggesting periodic pulsations of young stars due to mass growth during rapid accretion periods. 

Maser flares display significantly more rapid variability, with flux increasing several times over a short period. It is worth noting that there is no agreed-upon threshold of variability to define as flare among radio astronomers. Among the most remarkable and extensively studied flares are the methanol maser outbursts caused by accretion bursts in sources like S255IR$-$NIRS3 and G358.93$-$0.03 (\citealt{caratti2017}; \citealt{stecklum2021}). Monitoring efforts by single-dish radiotelescopes has proven to be an effective tool for identifying these extraordinary events (\citealt{fujisawa2015}; \citealt{Szymczak}; \citealt{macleod2018} and \citealt{burns2020}).

The causes of variability in 6.7\,GHz methanol masers continue to be a prominent area of research. \citet{caswell1995} proposed that maser variations are a result of changes in the gain path length caused by large-scale gas motions. In periodic 6.7\,GHz sources, a strong correlation between maser and infrared flux densities has been observed (e.g., \citealt{olech2020, olech2022}; \citealt{kobak2023}), indicating that the pumping rate plays a dominant role. A series of studies that combine single-dish time series with high angular resolution Very Large Baseline interferometry (VLBI) data have shown that most of the 6.7\,GHz methanol maser variability is caused by variations in pumping rates (e.g., \citealt{szymczak2014}; \citet{Moscadelli_2017}; \citealt{olech2019} and \citealt{durjasz2019}).

In this paper, we present the results of a flux density monitoring for 42 methanol maser sources, spanning up to 5 years. This program's initial goals and source list was firstly presented in \citet{aberfelds_shmeld_berzins_2017}. Our main motivation was to investigate maser variability and assess the capabilities of the Irbene facility for maser observations and highlighting the importance of maser monitoring. Previously mentioned extraordinary methanol maser outburst (S255IR$-$NIRS3 and G358.93$-$0.03 \citealt{fujisawa2015}; \citealt{Sugiyama_2019}) have been detected by medium-size radiotelescopes, very similar to those in Irbene.
The Irbene facility has a rich and unusual history. After the downfall of the Soviet Union, these military severance antennas were left in poor technical condition for two decades. However, between 2014 and 2016, the radiotelescopes underwent reconstruction, upgrades, and were equipped to modern standards, making scientific studies finally feasible with this instrument. As a relatively recent asset to the broader radio-astronomy community, several scientific programs have been realised. Including studies of Solar Corona activities, Active Galactic Nuclei (AGN) monitoring, the maser monitoring program presented in this study, monitoring of fast radio bursts (FRBs) under the framework of PRECISE project\footnote{\url{http://www.ira.inaf.it/precise/Home.html}}, and participation in the European VLBI Network (EVN)\footnote{The European VLBI Network is a joint facility of independent European, African, Asian, and North American radio astronomy institutes} (\citealt{Sukharev_2022}, \citealt{Younes_2022}, and \citealt{Ryabov_2023} as examples). 
Parts of monitoring results for separate methanol maser sources are reported in \citet{salii_2022}; \citet{aberfelds2021} and \citet{Aberfelds_2023a}. This is the first proper reporting of a complete set of our 6.7\,GHz methanol maser monitoring results.
    
\section{Observations}
\subsection{Instrumentation and methods}\label{Instrumentation and methods}
The 32 m and 16 m radio telescopes (RT--32 and RT--16, respectively) operated by Ventspils International Radio Astronomy Center (VIRAC), located in Irbene, Latvia were used for the monitoring of the 6668.519 MHz methanol maser transition. The flux density was sampled with irregular frequency, but typically, each source in our 42 source sample was observed once every 5 to 7 days. Sources exhibiting noticeably faster variability (e.g., more than a 33 percent flux difference for a spectral component between two observations) were selected for daily observations in order to delineate periods of enhanced activity. There are several up to thirty--day gaps in moitoring observations due to scheduling constraints and a few instances when both telescopes were not operational. Around 95 per cent of measurements were carried out on the RT--16.
    
Both telescopes are equipped with broad band 4.5 -- 8.8 GHz dual--polarisation receivers. The signal chains for both telescopes are identical. The wide-band feed-horns are optimised for each telescope's geometry, and both are calibrated and operated in the same manner. The RT--32 has a half-power beam-width (HPBW) of 6 arcmin but it suffers from relatively strong side lobes which may cause confusion, the HPBW of the RT--16 is 12 arcmin. The best practically achievable pointing accuracy is 30 arcsec for both telescopes. Estimated telescope efficiencies are ~0.4 and ~0.8 for RT-32 and RT–16, respectively. Surface accuracy (rms) of RT--32  primary mirror is around 3.2 mm and 635 $\mu$m for RT--16. Both telescopes use a noise diode with generated signal of (T\textsubscript{cal}) of 3.82 K. Elevation depended Degrees Per Flux Units (DPFU) at zenith are 0.086 K Jy$^{-1}$ and 0.046 K Jy$^{-1}$, for RT--32 and RT--16, respectively \citep{RT-16_performance}.
Both telescopes use fast Fourier transform (FFT) back-end spectrometers based on the Ettus Research USRP X300\footnote{https://www.ettus.com/all-products/x300-kit/} software-defined radio (SDR)  platform \citep{Bleiders}. Standard setup parameters for monitoring observations were 4096 channels per 1.5625 MHz band giving 0.017~km~s$^{-1}$ spectral resolution. Frequency switching mode \citep{Winkel} was used with four 15 sec long integration stages (first pair with noise diode off, last pair with noise diode on). Typical on--source integration times were 15 min averaging multiple foursome stages together. Stages that produced unsatisfactory results can be omitted from the averaging if needed.

Systems were regularly calibrated using stable continuum sources obtained from \citet{Perley_2013} and by stable maser sources like G32.744--0.076; G69.540-0.976 and G75.782+0.34 \citep{Szymczak}. For adequate pointing accuracy, we employed antenna Field System (FS) \citep{2000ivsg.conf...86H} pointing models, which are regularly calibrated by observing strong continuum sources. Under good observation conditions, the system temperatures were 28 K and 33 K for the RT--32 and RT--16, respectively.  Data processing was done using an in--house developed software package \textsc{maser-data-processing-suite (mdps)} described in detail by  \citet{Maser-Data-Processing-Suite}.  

\subsection{Source selection}    
Sources were selected from the Torun methanol source catalogue \citep{Toruncatalog} with the following criteria: declination above -10$^{\circ}$ and peak flux density above 3 Jy. A significant portion of our selected sources have been extensively studied in the past, enabling comparison of our results with those of previous research. Several sources were added later, as follow up observations in reaction to alerts of heightened maser activity from the Maser Monitoring Organization\footnote{https://www.masermonitoring.com/}.  The complete list of targets is given in Table \ref{table:sources}.  The source names are given according to their Galactic coordinates, their coordinates and systemic radial velocities (V\textsubscript{\it{lsr}}) are obtained from \citet{Toruncatalog} with the exception of G85.411+0.002 for which observation parameters were provided by private communication with the Maser Monitoring Organization.

\begin{table}
\caption{Methanol maser sources monitored at 6.7 GHz, under our monitoring program. }
\label{table:sources}      
\centering                                      
\begin{tabular}{l l l l }          
\hline                   
Source & RA(J2000) & Dec(J2000) & V\textsubscript{\it{lsr}}\\ 
        &(h m s) & ($^o$ ' ") & (km~s$^{-1}$) \\
\hline                                   
    G22.357+0.066 & 18 31 44.12 & -09 22 12.3 & 79.4  \\ 
    G24.33+0.14 & 18 35 08.09 & -07 35 03.6  & 112.0 \\      
    G25.709+0.044 & 18 38 03.15 & -06 24 14.9 & 95.5\\
    G25.64+1.05 & 18 34 21.99 & -05 59 38.6      & 41.0 \\ 
    G30.99-0.08 & 18 48 10.80 & -01 45 39.3      & 77.8  \\  
    G32.04+0.06 & 18 49 36.6 & -00 45 45.6 & 92.7 \\         
    G32.744-0.076 & 18 51 21.87 & -00 12 05.3      & 35.0 \\    
    G33.641-0.228 & 18 53 32.56 & 00 31 39.2      & 60.0 \\   
    G35.20-1.74 & 19 01 46.90 & 01 13 07.5      & 44.0 \\       
    G34.396+0.222 & 18 53 18.00 & 01 25 24.55      & 60.0 \\   
    G36.705+0.096 & 18 57 59.123 & 03 24 06.11      & 62.2 \\   
    G37.479-0.105 & 19 00 07.14 & 03 59 53.3 & 59.1 \\ 
    G37.43+01.51 & 18 54 14.23 & 04 41 41.1 & 41.3 \\          
    G37.55+0.20 & 18 59 09.986 & 04 12 15.6   & 85.0 \\           
    G43.149+0.013 & 19 10 11.05 & 09 05 20.4      & 15.0 \\    
    G43.796-0.12 & 19 11 54.016 & 09 35 49.46 & 40.0 \\      
    G45.071+0.132 & 19 13 22.129 & 10 50 53.11 & 57.8  \\  
    G49.04-1.08   & 19 25 22.30 & 13 47 20.1    & 37.1   \\
    G196.454-01.677 & 06 14 37.03 & 13 49 36.6 & 14.7 \\
    G49.490-0.388 & 19 23 43.96 & 14 30 35.0      & 57.9 \\  
    G192.60-0.05 & 06 12 54.02 & 17 59 23.3 & 6.5 \\          
    G189.030+0.784 & 06 08 40.67 & 21 31 06.9 & 9.6 \\        
    G59.783+0.065 & 19 43 11.25 & 23 44 03.3 & 19.5 \\        
    G69.540-0.976 & 20 10 09.074 & 31 31 35.95      & 7.5 \\  
    G174.20-0.08 & 05 30 48.01 & 33 47 54.6 & 3.5 \\
    G173.482+2.446 & 05 39 13.06 & 35 45 51.3      & -12.0 \\
    G73.06+1.80   & 20 08 10.20  & 35 59  23.7 & 6.1  \\    
    G75.782+0.34 & 20 21 44.20 & 37 26 36.7 & -0.5 \\        
    G78.122+3.633 & 20 14 25.88 & 41 13 36.87  & -6.5 \\     
    G81.88+0.78 & 20 38 36.45 & 42 37 36.1      & 5.5 \\      
    G188.95+0.89 & 06 08 53.34 & 42 37 36.1      & 10.5 \\     
    G85.411+0.002 & 20 54 13.689 & 44 54 07.686  & -29.4  \\  
    G90.92+1.49 & 21 09 12.60 & 50 01 02.9      & -69.2 \\    
    G94.602$-$1.796 & 21 39 58.26 & 50 14 20.9      & -43.0 \\  
    G123.066-6.309 & 00 52 24.20 & 56 33 43.2      & -31.0 \\   
    G111.26-0.77 & 23 16 10.00 & 59 55 31.3      & -37.0 \\     
    G111.542+0.777 & 23 13 45.36 & 61 28 10.6      & -56.2\\   
    G133.947+1.064 & 02 27 03.82 & 61 52 25.4      & -43.0 \\    
    G109.871+2.114 & 22 56 17.90 & 62 01 49.7      & -3.5 \\   
    G121.298+0.659 & 00 36 47.35 & 63 29 02.2      & -25.8 \\  
    G107.298+5.639 & 22 21 26.81 & 63 51 37.14    & -8.5 \\    
    G108.184+5.519 &  22 28 51.41 & 64 13 41.3      & -11.0 \\   
\hline                                             
\end{tabular}
\end{table}

\subsection{Error evaluation}\label{error_sub_sec}
As mentioned above, both instruments were calibrated and operated identically. Long-term pointing precision control suggest an average target offset of 2 arcmin for RT--16 and 1.1 arcmin for RT--32 corresponding to 92 per cent of normalized power. System temperature stability estimations suggest this contributes as a relatively minor (no more than 5 per cent) cause of amplitude error \citep{2020A&AT...32...23B}.
Individual measurement error was estimated as two times the noise signal standard deviation $\sigma$ (with average values of 1.9 and 0.9 Jy for RT--16 and RT--32, respectively) plus signal amplitude $A$ multiplied by relative error $\eta$  (ranging from 7 to 22 per cent) i.e. $2\sigma+A\eta$. Relative error was estimated from the comparison of multiple intrinsically low-variable source flux measurements with their corresponding mean flux densities. Observational data with clearly excessive noise levels, abnormally high system temperature or spectra with bad baseband slopes were discarded to maintain the integrity of the final averaged results. Overall, we conclude that absolute amplitude uncertainty is around 20 per cent.
To avoid falsely interpreting instrumental effects as source variability, we took several actions: observations were scheduled at as consistent as possible elevation angles for a given source, avoiding observations during strong winds and rain, and by conducting regular calibration sessions.

\subsection{Variability parameters}\label{Variability parameters}

From the obtained source spectra, the tracked flux densities of the components (as shown in light curves and used for variability analysis) are measured at their corresponding peak channels. In other words, each spectral line has a specific radial velocity, and we report the flux from the channel with the highest intensity. Doppler correction for local standard of rest frame was used to align measurements at different epochs. Since methanol as a paramagnetic molecule shows a only weak polarisation distribution \citep{gray_2012}, to increase the signal-to-noise (S/N) ratio for further analysis we used a flux value averaged over both polarisations. 

To characterise variability of a spectral feature, the same statistical tools e.g. variability index, were employed, as done by \citet{Szymczak} and \citet{Goedhart}. The variability index, originally introduced by \citet{Stetson_1996}, but reintroduced in a more usable form by \citet{Aller_2003} is: 
\begin{equation}
VI = \frac{(S_{max} -\sigma_{max})-(S_{min} +\sigma_{min})}{(S_{max} -\sigma_{max})+(S_{min} +\sigma_{min})}
\end{equation}
Here, $S_{max}$ and $S_{min}$ are highest and lowest flux density values, respectively, and $\sigma_{max}$ and $\sigma_{min}$ are absolute flux uncertainties in these measurements. $VI$ values range from 0 for non--variable behaviour to 1 for strongly variable lines. Occasionally value can be negative, indicating that measurement uncertainty is larger than the measurement value itself. It should be noted that $VI$ is also significantly affected by outliers, as, in essence, only two measurement values are used. 

Introduced by \citet{Aller_2003}, the  fluctuation index, in our opinion, provides a better statistical tool to characterise low signal-to-noise spectral features and low-amplitude variability. It also has the advantage of a lower sensitivity to outliers. Fluctuation index measures spread versus the mean flux density value. 
\begin{equation}
FI = \left[ \frac{N}{\sum_{i=1}^{N}\sigma_{i}^2 }\left(\frac{\sum_{i=1}^{N}S_{i}^2\sigma_{i}^2 - \overline{S}\sum_{i=1}^{N}S_{i}\sigma_{i}^2}{N-1}-1\right) \right]^{0.5}/\overline{S}
\end{equation}

Here, $S_{i}$ is an individual flux density measured in specific epoch $i$, $\overline{S}$ is the average flux density over the time interval under evaluation, $\sigma_{i}$ is the individual measurement uncertainty and $N$ is the number of measurements.

Following \citet{Szymczak}, we also included the $\chi^2_{r}$ parameter as our final metric for evaluating variability: 

\begin{equation}
\chi^2_{r}=\frac{1}{N-1}\sum_{i=1}^{N}\left(\frac{S_{i}-\overline{S}}{\sigma_{i}}\right)^2
\label{chi2red_formula}
\end{equation}
For intrinsically non-variable sources this parameter will be close to unity.

\subsection{Wavelet analyses}\label{Wavelet analyses}
Wavelet spectral analysis has emerged as a new and powerful tool for studying monitoring data. We used open source script \textsc{pycwt} \citep{Sebastian2023} which follows the approach suggested by \citet{Torrence_1998} and \citet{Liu_2007} for time series analyses. Time series input, script parameters and visualisation suggestions are made following \textsc{pycwt} documentation and examples there. Wavelet transformation requires input data to be sampled at regular intervals which is not typically the case for maser monitoring observations, thus data re-sampling is necessary. For this purpose a linear interpolation function \textsc{interp1d} from \textsc{scipy} \citep{2020SciPy-NMeth} was used. These re-sampled data sets were normalised by standard deviation and de-trended by linear a function. Global Wavelet Spectra in combination with Fast Fourier Transformation (FFT) using re-sampled and de-trended data, and Lomb-Scargle periodogram \citep{scargle1982} using the original data set, were used to search for periodicity signals in the maser time series. 

Wavelet analysis plots (e.g. Figure \ref{g107_7.4_wavelet}.) show: panel a), obtained light curve denoted with red dots, overlaid with linearly interpolated time series, shown as a black line.
Panel b) shows the Wavelet Power spectrum of time series in which the x and y axes represent time and period, and colour indicates the strength or power of the signal's frequency components at that particular time and frequency \citep{Torrence_1998}.
More intense colours indicate higher power or stronger presence of periodicity at a particular frequency. The crossed out area represents combinations of time and signal periods which are affected by boundary effects i.e. the cone of influence. Panel c) shows the global wavelet power spectrum, which represents the overall presence of a particular frequency in the entire time span, overlaid with the FFT spectra and Lomb-Scargle periodogram. Here also a 95 per cent confidence level for significant frequency components in the time series are presented. Note that the y axes of b) and c) panels have logarithmic scale and trivial relations between frequency and period are used. We consider a source to have a statistically significant period in their time series if global wavelet power spectra, FFT and Lomb-Scargle periodogram display agreement and show a clear signal over 95 per cent confidence.

\section{Results}
We have conducted extensive flux monitoring for forty-two 6.7 GHz methanol masers, spanning up to five years.
During the observing program, we noticed various maser variability types. The summary of source variability by types, described below, are presented in Table \ref{variability_type}.
 
\textbf{Low--variable.} We categorise a maser source as having low variability when the variability indexes for all its spectral components are below 0.25. In this category are 21 per cent sources from our sample. Note, G133.947+1.064/W3(OH) two low-intensity lines at -41.8 and -42.2~km~s$^{-1}$, which, due to significant blending, are excluded from our analyses, despite their variability. 

\textbf{Moderately variable.} The variability index of the most variable component falls between 0.26 and 0.5. This variability type constitutes 26 percent of all sources.

\textbf{Highly variable.} A source is classified as highly variable if the variability index of at least one of its spectral components is above 0.5. This criterion was met for 55 percent of the sources in our study. 

\textbf{Correlated fluctuations.} In this category, sources with two or more spectral features whose variability time series show a high (>0.7) positive correlation coefficient. Only sources with notable variability (variability index >0.25) are considered, we found that 19 per cent of sources have this variability type.

\textbf{Anti--correlated fluctuations.} Similar to the above but for spectral features exhibiting high (<-0.7) negative correlation coefficient. For example: in G33.641-0.228 the feature pair at 59.6 and 59.3~km~s$^{-1}$ are anti-correlated with features at 63.2 and 62.7~km~s$^{-1}$; in G109.871+2.114/Cepheus A, the feature at -2.1~km~s$^{-1}$ is strongly anti - correlated with features at -3.7, -4.0 and -4.7~km~s$^{-1}$. Only 7 per cent of our sources have this type of variability.

\textbf{Velocity drifting.}  Refers to a continuous change in the radial velocity of a spectral line, with a drift of at least several spectral channels (>0.07~km~s$^{-1}$) of the observed maximum velocity of the line. In our sample, only a single source (G121.298+0.659 see Figure \ref{G121_dynamic}.), or 2 percent, exhibited this kind of behaviour.

\textbf{Cyclical variability.} This category comprises sources with statistically probable periodical variability. In this work we used wavelet analyses and the Lomb-Scargle periodogram to identify statistical probable (wavelet spectrum, FFT and Lomb-Scargle all agrees and show clear signal over 95 percent confidence, see section \ref{Wavelet analyses}) periodic signals.  In this category are 19 per cent of sources from our sample. Parameters of sources with this variability type are summarised in Table \ref{quasi-periodic sources}. We show: the approximate period of variability, observed amount of flux change cycles, flux variation range (typical minimum and maximum flux levels for the spectral line), percentage of relative flux increase from the minimum and we provide a bibliographic references for previously identified periodicity in known sources. 

\textbf{Raising.} A source is considered to be rising if its flux density steadily increases by more than 50 percent over a period of 100 days. We observed this rising flux trend in 7 percent of the sources in our sample. 

\textbf{Falling.} Similarly, if a spectral feature steadily loses flux by more than 50 percent in less than 100 days, we classify it as a falling flux source. We observed this type of flux decrease in 10 percent of sources. Importantly, there is a significant overlap between the rising and falling flux sources, where the flux density of the same spectral components alternately and steadily increases or decreases. 

\textbf{Flaring.} As mentioned before a flare is a rapid increase in maser flux density. We propose identifying a flux increase as a flare when the flux density exceeds above a rolling mean flux and five times the statistical standard deviation of the flux within that time span. More details on our flare definition are given in Section 4. If the period of increased flux extends beyond 25 days, we further classify it as a \textbf{long flare}. Conversely, if the flux returns to its usual levels in less than 25 days, we categorize it as a \textbf{short flare}. After a flare, we see that the flux usually returns to its pre-flare intensity, but the rise and decline profiles of flare are often source-distinctive, as noted by \citet{Szymczak}. We detected flaring activity in 10 percent of the sources in our sample.

\null 

The mean values of variability parameters in our sample are 0.33 for the variability index, 0.54 for the fluctuation index and 4.66 for the $\chi^2_{r}$ parameter.
The highest variability index in our sample was found to be 0.99 for the periodically flaring G107.298+5.63. This maser source also has the highest fluctuation index and $\chi^2_{r}$ parameter ($FI$ = 6.76 and $\chi^2_{r}$ = 307.6). Overall, all spectral components of G107.298+5.639 have highly variable parameters, indicating that it is the most variable source in our sample.  
Most of the identified variable lines with a variability index above 0.5 are found in low flux (S\textsubscript{P} $\leqslant$ 30 Jy) maser features. Similar relations were found with the fluctuation index and $\chi^2_{r}$ parameter, which aligns with the findings of \cite{Goedhart} and \cite{Szymczak}.
There is a clear positive correlation trend between the variability index, fluctuation index, and the $\chi^2_{r}$ parameter. In total, we found 54 out of 190 spectral features to be highly variable ($VI$ > 0.5), belonging to 24 different sources, with 15 of them being faint features.


\begin{table}
\centering
\caption{Detected variability types} 
\begin{tabular}{l c}
\hline
Source  & Variability type \\
\hline
G22.357+0.066 &   Highly variable; Correlated fluctuations; Cyclical\\ 
G24.33+0.14 & Highly variable; Long flare\\     
G25.709+0.044 & Low--variable \\
G25.64+1.05 &  Moderately variable \\ 
G30.99-0.08 &  Highly variable \\  
G32.04+0.06 &  Moderately variable; Cyclical \\         
G32.744-0.076 &  Low--variable \\    
G33.641-0.228 &  Highly variable; Correlated fluctuations; Long flare\\
              & Anti--correlated fluctuations; Short flare; Cyclical \\   
G35.20-1.74 &  Highly variable; Correlated fluctuations; Falling \\       
G34.396+0.222 &  Moderately variable \\   
G36.705+0.096 &  Low--variable \\   
G37.479-0.105 &  Low--variable \\ 
G37.43+01.51 &  Low--variable \\          
G37.55+0.20 & Moderately variable; Correlated fluctuations; Cyclical \\                   
G43.149+0.013 &  Low--variable \\    
G43.796-0.12 &  Low--variable; Correlated fluctuations \\      
G45.071+0.132 &   Low--variable \\  
G49.04-1.08   &   Highly variable \\
G196.454-01.677 &  Highly variable; Correlated fluctuations; Cyclical\\
G49.490-0.388 &  Moderately variable \\  
G192.60-0.05 &  Highly variable; Cyclical \\          
G189.030+0.784 &  Highly variable \\        
G59.783+0.065 &  Highly variable; Raising; Falling; \\
              & Anti--correlated fluctuations \\        
G69.540-0.976 &  Moderately variable \\  
G174.20-0.08 &  Highly variable; Falling \\
G173.482+2.446 &  Moderately variable \\
G73.06+1.80   &   Highly variable; Cyclical \\    
G75.782+0.34 &  Highly variable\\        
G78.122+3.633 &  Highly variable; Falling; \\
& Raising; Correlated fluctuations \\     
G81.88+0.78 &  Moderately variable \\      
G188.95+0.89 &  Highly variable \\     
G85.411+0.002 &  Moderately variable; Long flare \\  
G90.92+1.49 &  Moderately variable \\    
G94.602$-$1.796 &  Highly variable \\  
G123.066-6.309 & Highly variable \\   
G111.26-0.77 &  Highly variable \\     
G111.542+0.777 & Moderately variable \\   
G133.947+1.064 &  Low--variable \\    
G109.871+2.114 &  Highly variable; Correlated fluctuations; \\
               &  Anti--correlated fluctuations \\   
G121.298+0.659 &  Highly variable; Raising; Falling; Velocity drifting \\  
G107.298+5.639 &  Highly variable; Short flare; Cyclical \\    
G108.184+5.519 &  Highly variable \\ 
\hline
\end{tabular}
\label{variability_type}
\end{table}

\begin{table*}
\centering
\caption{Parameters of periodic (cyclical) sources. The variation range displays the typical minimum and maximum flux densities within a cycle. In the case of several spectral lines following the same periodic pattern, we utilized the parameters of the most intense lines. The relative increase indicates the percentage by which the minimum amplitude increased during the cycle.} 
\begin{tabular}{l c c c c c}
\hline
Source & Approximate Period & Observed number of & Variation &  Relative & Reference\\
       & days               & cycles             & range (Jy) & Increase (\%) & \\
\hline
G22.357+0.066   & 170  & 9.2 & 11-37   & 240  & \citet{Szymczak_2011}\\
G32.04+0.06     & 57   & 11  & 108-200 & 90  & This work\\
G33.641-0.228   & 500  & 4.4 & 5-55    & 100   & \citet{olech2019}\\
G33.641-0.228   & 115  & 7 & 40-95    & 130   & This work\\
G37.55+0.20     & 250  & 5.2 & 3-13    & 330  & \citet{Araya_2010}\\
G73.06+1.80     & 123  & 5   & 4-14    & 250  & \citet{Szymczak_2015}\\
G192.60-0.05    & 235  & 5.5 & 60-120  & 100    & This work\\
G196.454-01.67  & 110  & 7   & 5-32    &  600   & \citet{Szymczak}\\
G107.298+5.63   & 34.4 & 43  & *-250   & $>2.5\times10^{4}$ & \citet{Szymczak_2016}\\
\hline
\multicolumn{2}{l} {* Flux regularly fell bellow our detection level. }
\end{tabular} 
\label{quasi-periodic sources}
\end{table*}

\begin{figure*}
\centering
\includegraphics[width=0.85\paperwidth, height=0.22\textheight]{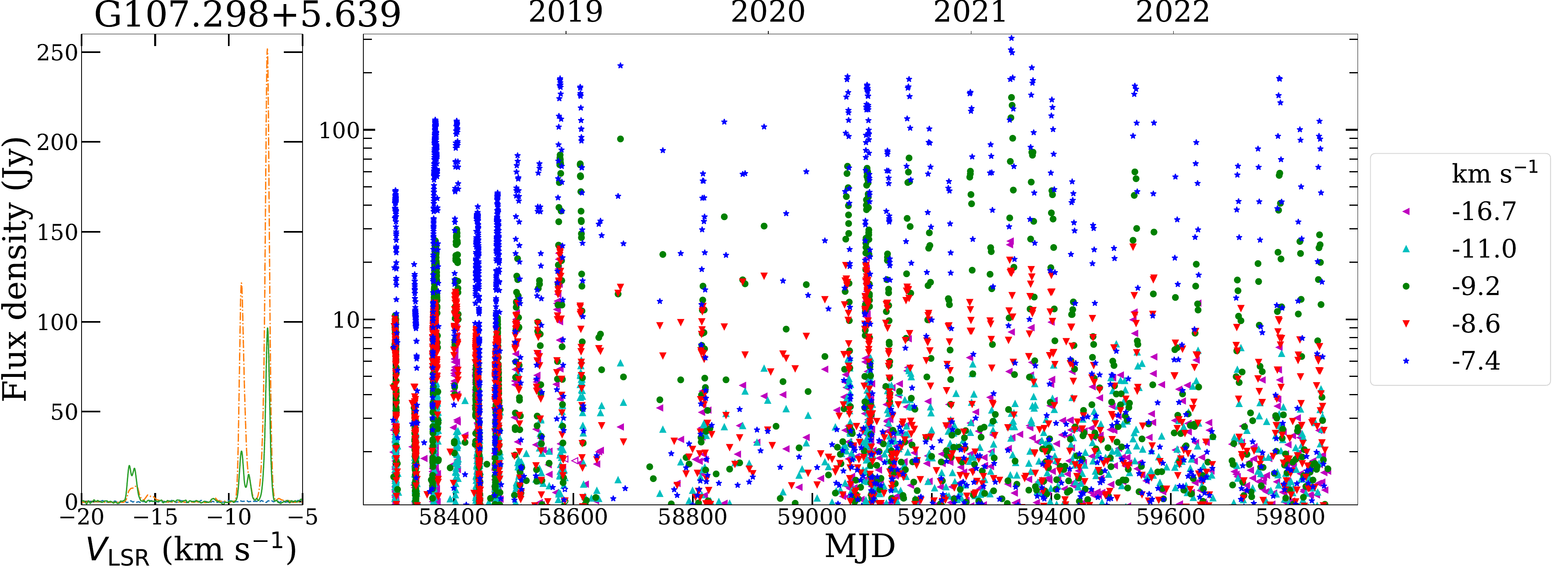}
\includegraphics[width=0.85\paperwidth, height=0.22\textheight]{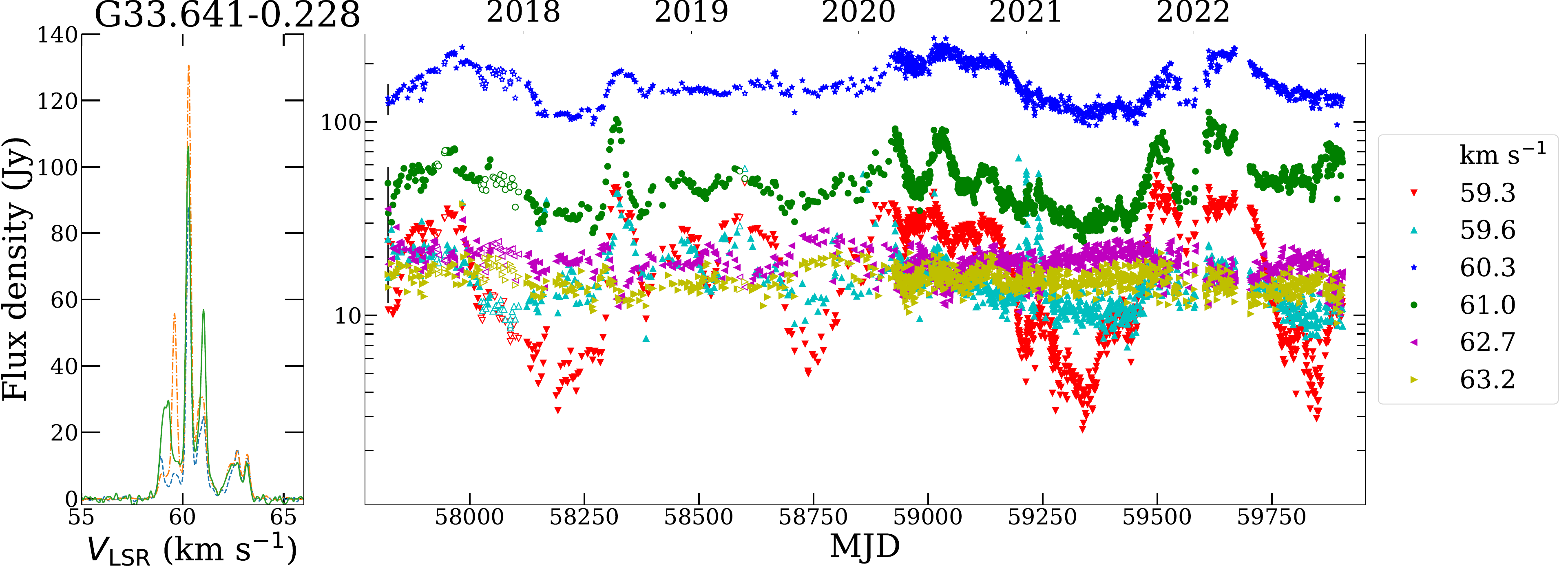}
\includegraphics[width=0.85\paperwidth, height=0.22\textheight]{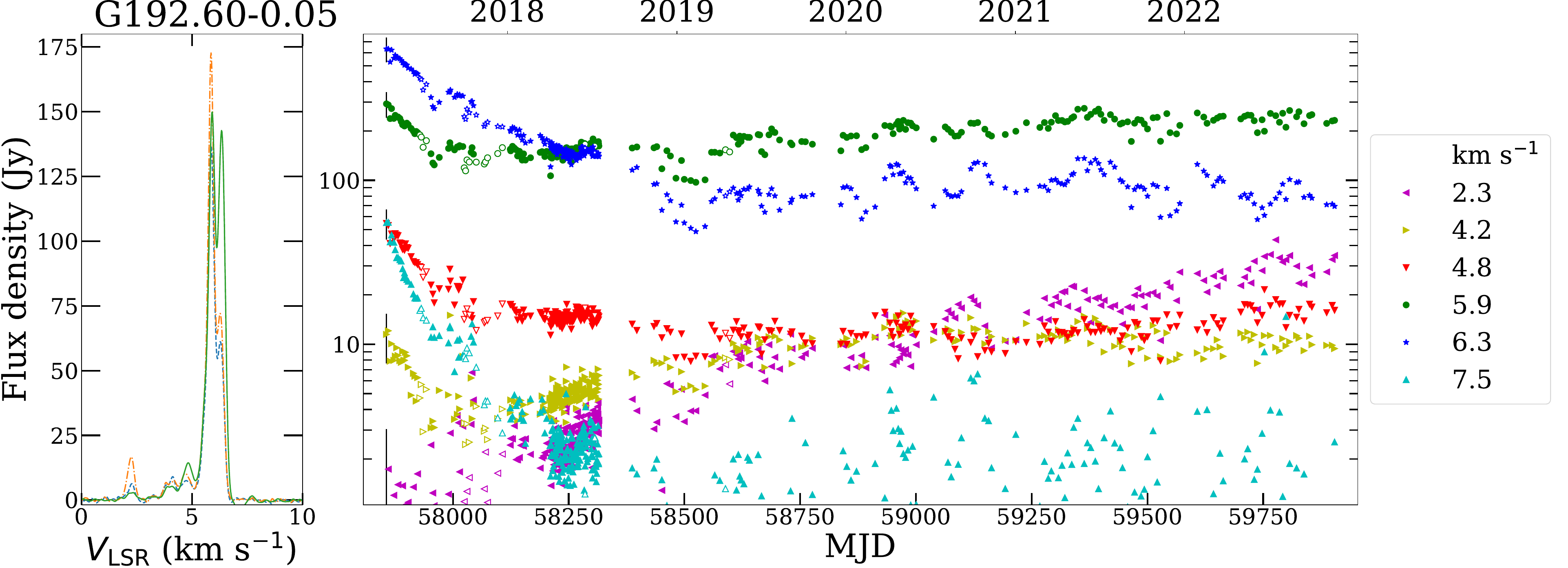}
\caption{Spectra and light curves of the 6.7\,GHz maser emission in the selected objects. Left-hand panels: maximum (orange), minimum (blue) and some average (green) spectra that, we think, best represents usual flux densities. Right-hand panels: light curves of main features. The filled and empty symbols represent the data from the 16-m and 32-m telescopes, respectively. Typical measurement uncertainty is shown by the bar for the first data point, calculated as stated in section \ref{error_sub_sec}. Note logarithmic scale. } 
\label{Three_sourec_mon}
\end{figure*}

Time series for selected sources whose high variability and interesting trends merit additional discussion are presented in Figure \ref{Three_sourec_mon}. Variability parameters for these three sources are summarised in Table \ref{variability parametrs}, and for all monitored sources in our sample in Table \ref{full_var_table}.  Variability analyses for G78.122$+$3.633, G90.925$+$1.486 and G94.602$-$1.796 are presented in \citet{Aberfelds_2023a}.

\begin{table}
\centering
\caption{Statistics of the maser line variability time-series. Here,  V\textsubscript{P}(km s$^{-1}$) is a spectral feature's velocity at LSR, S\textsubscript{P}(Jy) is mean flux density value for given spectral component, $VI$ and $FI$ -- variability and  fluctuation indexes and $\chi^2_{r}$ parameter. MJD\textsubscript{s} is the start time of observations in Modified Julian Date, $T\textsubscript{s}$ is the total time-span of observation in years, $N$ - number of observations.} 
\begin{tabular}{l c c c c}
\hline
V\textsubscript{P}(km~s$^{-1}$) & S\textsubscript{P}(Jy) & $VI$ & $FI$ & $\chi^2_{r}$\\
\hline
\multicolumn{5}{l}{G107.298+5.639 (MJD\textsubscript{s}=58298, $T\textsubscript{s}$= 4.285\,y, $N$=1823)}\\
-16.7 &  1.71 & 0.91 & 3.09 & 1.96\\
-11.0 &  1.53 & 0.61 & 0.81 & 1.10\\
-9.2 &  7.46 & 0.98 & 6.76 & 23.56\\
-8.6 &  4.30 & 0.88 & 1.76 & 8.21\\
-7.4 &  27.04 & 0.99 & 4.81 & 307.60\\
\multicolumn{5}{l}{G33.641-0.228 (MJD\textsubscript{s}=57821, $T\textsubscript{s}$= 5.707\,y, $N$=1136)}\\
59.3 &  18.78 & 0.83 & 1.05 & 17.86\\
59.6 &  15.40 & 0.73 & 0.77 & 2.18\\
60.3 &  161.73 & 0.21 & 0.50 & 2.66\\
61.0 &  49.98 & 0.49 & 0.65 & 4.56\\
62.7 &  18.39 & 0.32 & 0.18 & 0.56\\
63.2 &  15.21 & 0.39 & 0.16 & 0.28\\
\multicolumn{5}{l}{G192.60-0.05 (MJD\textsubscript{s}=57856, $T\textsubscript{s}$= 5.476\,y, $N$=368)}\\
2.3 &  7.47 & 0.92 & 2.52 & 8.66\\
4.2 &  7.19 & 0.67 & 0.78 & 1.68\\
4.8 &  15.75 & 0.63 & 0.74 & 2.52\\
5.9 &  176.91 & 0.37 & 0.52 & 3.07\\
6.3 &  161.27 & 0.81 & 1.59 & 17.11\\
7.5 &  4.48 & 0.97 & 4.24 & 6.27\\
\hline
\end{tabular}

Variability parameters for all sources are presented in Table \ref{full_var_table}.

\label{variability parametrs}
\end{table}

\subsection{G107.298+5.639} 
\begin{figure*}
\centering
\includegraphics[width=0.8\paperwidth]{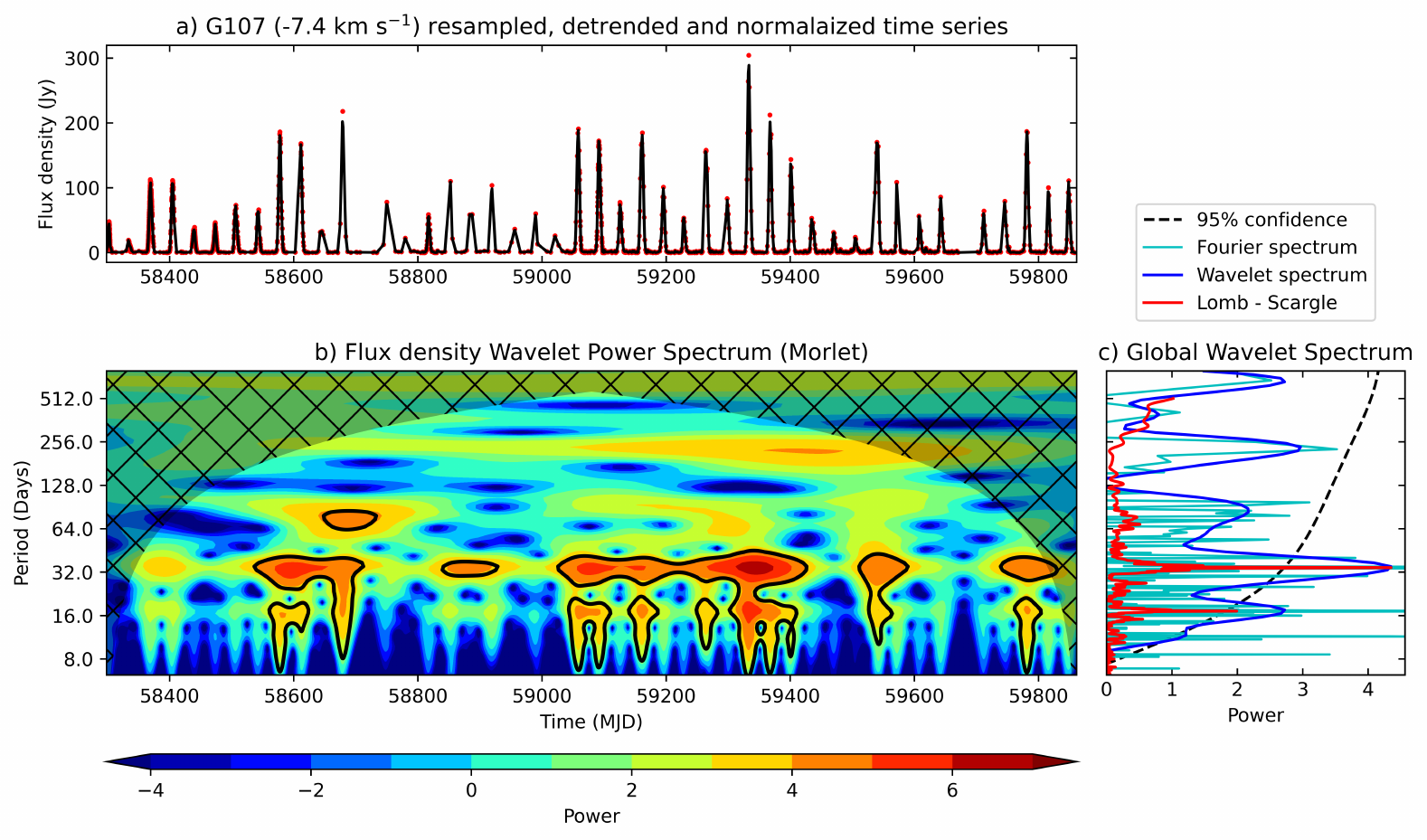}
\caption{a) Original G107.298+5.639 -7.4~km~s$^{-1}$ data denoted with red dots overlaid with regularly re-sampled, de-trended and normalised time series black line. b) Morlet wavelet power spectrum. c) Global wavelet power spectrum with significance level, additionally Fourier spectrum and Lomb - Scargle periodigram is also shown.  } 
\label{g107_7.4_wavelet}
\end{figure*}

\begin{figure}
\centering
\includegraphics[scale=0.38]{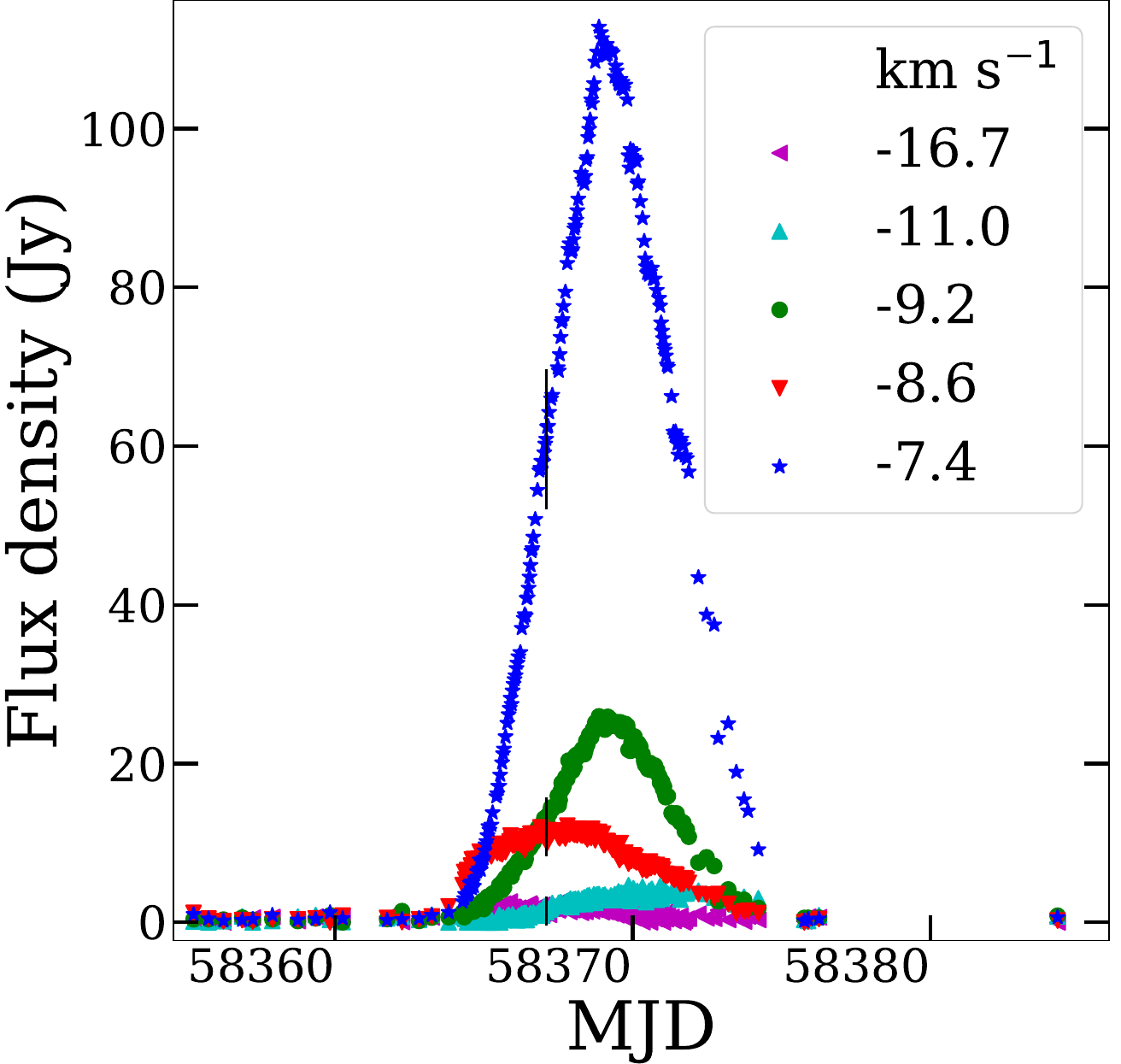}
\caption{Example of one G107.298+5.639 flare, demonstrating time lags between spectral components, a result made possible by the high--cadence observations. During the flare, flux density was sampled every 15 minutes. } 
\label{g107_1flare}
\end{figure}

G107.298+5.639 is a very variable source with periodical flares, between which maser emission is under our detection level (Figure \ref{Three_sourec_mon}. top). In total, we have observed 43 flares of all spectral features during the 4.3 years, and according to their expected period of 34.4 days \citep{Szymczak_2016}, only 2 flares are missing in our time series due to gaps. The rate of observations was increased from once a week to several observations per hour. The overall observed variability is consistent with the results of \citet{Szymczak_2016}. The spectral feature at -7.4~km~s$^{-1}$ is always the most intense, and its peak intensity may vary up to a factor of ten from flare to flare. There are noticeable time lags, for example the spectral feature at -8.6~km~s$^{-1}$ peaks around 2.2 days before the  -7.4 and -9.2~km~s$^{-1}$ features (see Figure \ref{g107_1flare}.) and these delays between components' peak flux moments are consistent flare to flare. Wavelet analyses of the -7.4~km~s$^{-1}$ spectral feature's time series are presented in Figure \ref{g107_7.4_wavelet}. Where panel a) shows original data (red dots) overlaid with re-sampled, de-trended and normalised time series (black line). Panel b) shows a Morlet Wavelet power spectrum and panel c) shows the Global Wavelet powers spectrum overlaid with the FFT spectrum and Lomb-Scargle periodogram. Figure \ref{g107_7.4_wavelet} clearly shows a significant 35 day periodic signal, which is consistent with the results of \citet{Szymczak_2016}. Overall the wavelet analyses method produces expected results.

\subsection{G33.641-0.22} 
\begin{figure*}
\centering
\includegraphics[width=0.8\paperwidth]{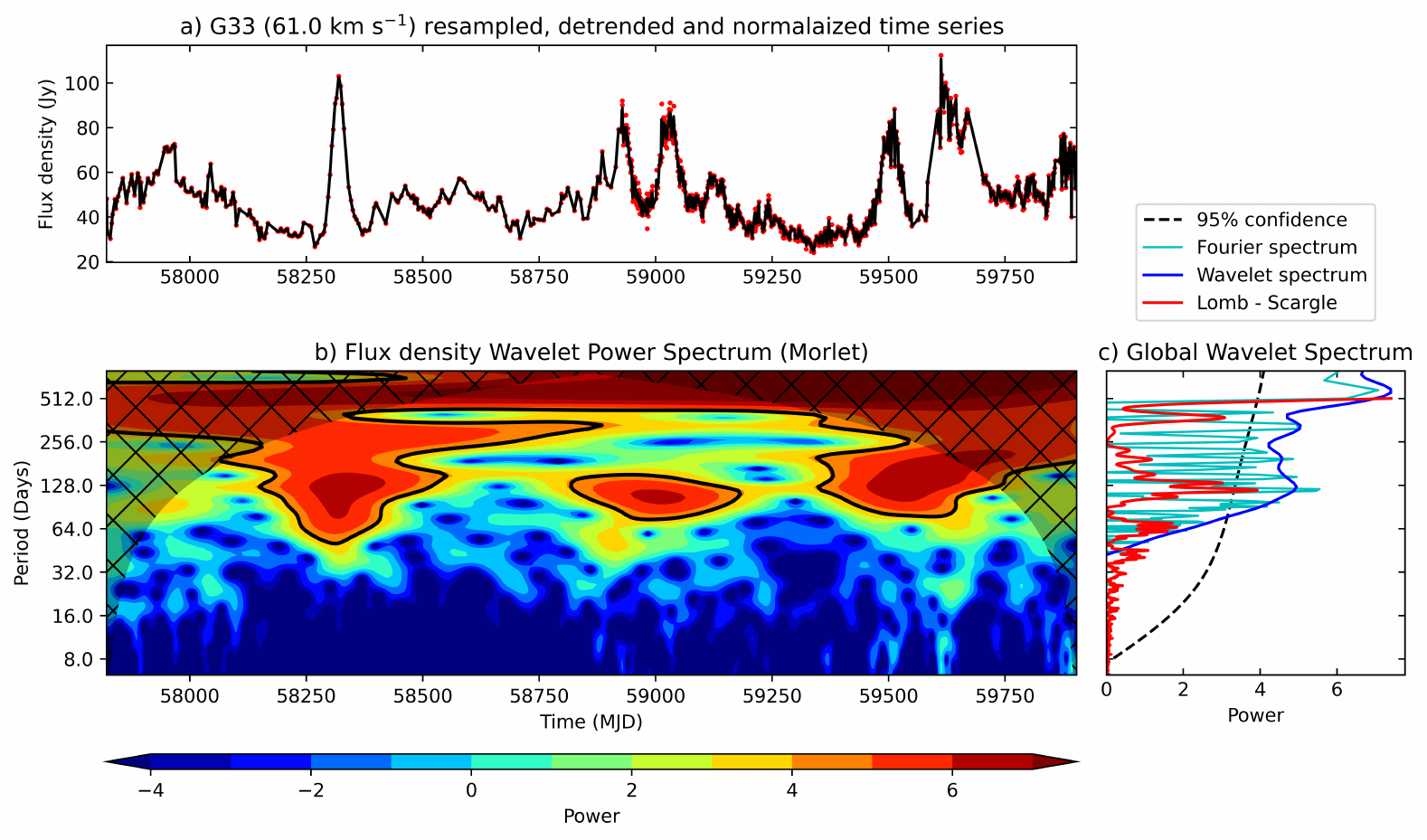}
\caption{Same as Figure \ref{g107_7.4_wavelet}., but for the G33.641-0.22 61.0~km~s$^{-1}$ line.} 
\label{g33_61_wavelet}
\end{figure*}

At the beginning, our target was to observe G33.641-0.22 once per every 4 days but, starting from 58929 MJD (March 2020), we switched to daily observations in response to it's high variability. Features at 59.3 and 59.6~km~s$^{-1}$ are strongly correlated and simultaneously anti-correlated with features at 62.7 and 63.2~km~s$^{-1}$. During the observation campaign we detected several flicker-like bursts similar to those previously reported by \citet{Fujisawa_2014}. According to our observations, new flares happened at  58153, 58167, 58601, 58895, 59015, 59103, 59170, 59216, 59241, 59264 MJD. Two additional bursts of feature at 61.0~km~s$^{-1}$ were detected, from 58270 to 58371 (Jun to September 2018) and 59460 to 59550 MJD (September to December 2021). Figure \ref{g33_61_wavelet} highlights the regularities found in the time series. There is a strong indication of long-duration period of 500 days which is in agreement with the findings of \citet{olech2019}. Additionally, there is a tentative detection of a shorter, approximately 115-day, periodic fluctuation, which coexists with other variability patterns. Note that these shorter period fluctuations are noticeable (have high power spectra) only in three periods: from 58150 to 58550 MJD (February 2018 to March 2019); from 58815 to 59180 MJD (full year from Decembers 2019 to 2020) and from 59350 to 59750 MJD (May 2021 to Jun 2022), coinciding with high flux density phases of this component (see Figure \ref{g33_61_wavelet}.). 

\subsection{G192.60-0.05} 
\begin{figure*}
\centering
\includegraphics[width=0.8\paperwidth]{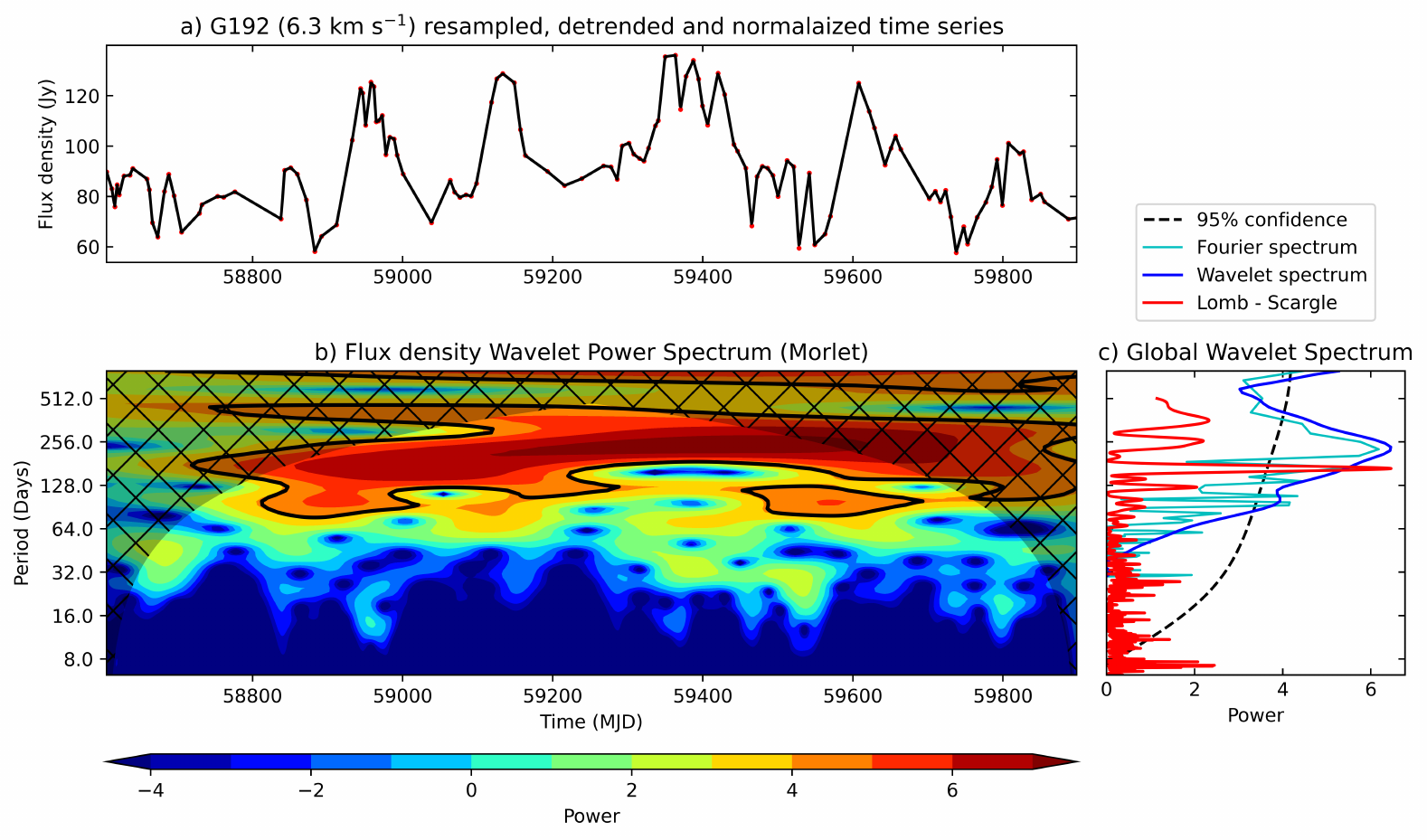}
\caption{Same as Figure \ref{g107_7.4_wavelet}., but for the G192.60-0.05 6.3~km~s$^{-1}$ line. Note, only data obtained alter 58600 (March 2018) MJD were used.  } 
\label{s255_6.3_wavelet}
\end{figure*}

G192.60-0.05 is also know as S255IR. This source was extensively studied during its massive flare event (\citealt{fujisawa2015} and \citealt{Moscadelli_2017}). Our methanol maser monitoring program started monitoring in February 2017 when the S255IR maser flare was already in its declining stage. Spectral features with high flux density (5.9 and 6.3~km~s$^{-1}$) continued to dim until 58325 MJD (July 2018), by a factor of 6, before becoming mostly stable, around 58600 MJD (April 2019). Starting from around 58600 MJD, the 6.3~km~s$^{-1}$ line may exhibit semi--regular fluctuations (Figure \ref{s255_6.3_wavelet}.). Notably, the wavelet power spectra indicates periodic fluctuations with increasing duration which may explain why Lomb - Scargle and other the methods do not agree well for the most probable length of periodicity (170 and 235 days respectively). Other spectral components do not show such fluctuations. The spectral line at 7.6~km~s$^{-1}$ faded down to our measurement noise level at around 58200 MJD (March 2018). Around 58000 MJD (September 2017), a new spectral feature at 2.3~km~s$^{-1}$ appeared.

\section{DISCUSSION}
\subsection{Comparison with relevant studies}
The variability of 6.7 GHz methanol masers has been a focal point in multiple studies. However, these studies often focus on one or a few sources (e.g. \citealt{Araya_2010}; \citealt{Szymczak_2016}; \citealt{olech2019})  or are large surveys that observe each source only a handful of times (e.g. \citealt{caswell1995} and \citealt{multibeam_catalog}). As previously mentioned, the two most significant long-term studies on 6.7 GHz methanol maser variability, in our opinion, have been conducted by \citet{Goedhart} and \citet{Szymczak}.

In our opinion, only these studies (\citealt{Goedhart}; \citealt{Yonekura2016}; \citealt{Szymczak}) with their corresponding methanol maser programs are reasonably comparable to ours based on both the large number of target sources and the comparatively high average observation cadence. A comparison with these studies is summarised in Table \ref{Comparison_with_other_studies}. There is practically no time overlap between all four studies, although it is likely that all teams are continuing 6.7 GHz methanol maser monitoring programs, and, in the future, direct time series comparisons and combinations will be possible. A high percentage of our sample sources were also monitored by \citet{Szymczak} and \citet{Yonekura2016}, with \citet{Goedhart} having a comparably smaller overlap percentage due to Hartebeesthoek radio observatory being located far South.

Our methanol maser monitoring program has a comparable average observation cadence to \citep{Yonekura2016} and an observation cadence 1.5 to 6 times more frequent than \citet{Goedhart} and \citet{Szymczak}. There is a notable difference in the percentage of detected highly-variable spectral components which have a variability index above 0.5 (see Table \ref{Comparison_with_other_studies}, eighth line), but this difference can be explained by selection effects; we favoured high flux density sources, which tend to be less variable, according to \citet{Szymczak}. Concluding, the quantitative characteristics of maser variability are in agreement with \citet{Goedhart} and \citet{Szymczak}, and our work and theirs are complementary.

\begin{table*}
\centering
\caption{Comparison with the studies by \citet{Goedhart}; \citet{Szymczak} and \citet{Sugiyama2019}.} 
\begin{tabular}{l c c c c}
\hline
 & \citet{Goedhart} & \citet{Szymczak} & \citet{Sugiyama2019} & This work\\
\hline
Start of monitoring program & Jan 1999 & Jun 2009 & Dec 2012 & Mar 2017\\
End of the examined period  & Mar 2003 & Feb 2013  & Nov 2016 & Oct 2022\\
Duration of monitoring (yr) & 4.2 & 3.7 & 3.9 & 5.3\\
Number of sources           & 54 & 137 & 442 & 42 \\
Percentage of sample overlap to ours & 24 & 90 & 97 & -\\
Average observation cadence per month & 2-4 & 1 & 6-8 & 6.5\\
Percentage of periodic sources & 13 & 7  & 10 & 19 \\
Percentage of high variable features & 55 & 54 & * & 28 \\
Percentage of low-variable sources & 19 & 21 & * & 21\\
\hline
\end{tabular}

* Unknown value

\label{Comparison_with_other_studies}
\end{table*}

\subsection{Instrument performance}
We the utilised ample availability of telescope time to obtain densely sampled time series data for particularly intriguing maser sources. Our high cadence even permitted obtaining four flux density measurements in one hour for a full-week (Figure \ref{g107_1flare}.), which maybe challenging or impossible for many other radio observatories.
As mentioned before, the vast majority (95 per cent) of the observations where made with the RT--16. Although telescope parameters are good for a 16 meter antenna, it cannot mach Torun's 32 meter telescope \citep{Szymczak} in sensitivity. Our monitoring results for sources with low fluxes are more affected by scattering caused by a lower signal to noise ratio. 

Our calibration error is comparably high, with relative amplitude precision around 20 per cent. In our view, these variations are not attributed to major flaws in the employed methods or instruments, but rather stems from specific challenges posed by seasonal weather conditions. Irbene is situated a mere 4.2 km from the Baltic Sea, causing rapid fluctuations in telescope gain due to changing weather patterns. The winter season presents the most dynamic and demanding conditions, as the Baltic Sea remains ice-free, and prevailing winds come from the sea, resulting in nearly 100 percent humidity for several months. These periods of high scattering can be noticed in all of the obtained time series. Inability to rapidly and accurately compensate for a changing signal gain is the main factor of high amplitude errors. Due to sporadic results of gain measurements in bad weather conditions, average values were used. Data suffering large deviations from expected gain values were discarded.

\subsection{Source variability}
No new types of variability were discovered in our observation sample; however, it is worth noting that several sources exhibited changes in their typical variability behaviour. Even sources that were previously considered stable can quickly become variable (e.g., G78.122+3.633/IRAS 20126+410 after a 500-day quiescence period, as shown in \citealt{Aberfelds_2023a})  and, conversely, highly variable sources (like G192.60-0.05/S255IR see Figure \ref{Three_sourec_mon}, bottom) can transition to a low-variability state. Notably, the three components of G59.783+0.065 (Figure \ref{Appendix_time_series}, bottom) became significantly variable after a year-long period of low variability, displaying distinct variability profiles. It is important to emphasise that the variability changes discussed in this paragraph are not related to flares.

\subsection{Periodic masers}

Eight sources from our sample (19 percent) exhibit noticeable periodic (cyclic) variations in flux density, but most of them have been previously identified by different research groups  (see summary in Table \ref{quasi-periodic sources}.). Our results for the G192.60-0.05 6.3~km~s$^{-1}$ line (Figure \ref{Three_sourec_mon}, bottom and Figure \ref{s255_6.3_wavelet}.) and G32.04+0.06 92.7~km~s$^{-1}$ line (Figure \ref{Appendix_time_series}, middle and  Figure \ref{G32p04_wavelet}.), and G33.641-0.228 61.0~km~s$^{-1}$ line (Figure \ref{Three_sourec_mon}. middle and Figure \ref{g33_61_wavelet}.), may indicate regularities (periods), previously not reported in the literature. Notably, these potential periodic signals arise from individual spectral features, which is somewhat rarer than the periodic variability of an entire source's spectra. Additionally, their relative flux increase (percentage by which minimal amplitude increased during cycle) are comparatively lower compared to already known periodic sources (see Table \ref{quasi-periodic sources}.). This highlights that a high observation cadence combined with methods like wavelet analysis, FFT, and the Lomb-Scargle periodogram may enable the detection of period signals with lower relative flux increases compared to already known periods reported in the literature regarding 6.7 GHz methanol maser variability.  

The longest cycle of a periodic signal we observed was 500 days (G33.641-0.228). Results of even longer-period variability in G33.641-0.228; G196.454-01.67 and G188.95+0.89/S252 were previously noted in the literature (\citealt{Goedhart}; \citealt{olech2019}); however, we didn't see concrete evidence of these during our observations. The shortest cyclic behaviour was 34.4 days (Figures \ref{Three_sourec_mon}, top and \ref{g107_7.4_wavelet}.) for G107.298+5.63, which is a well--studied source \citep{Szymczak_2016}. 
We would like to add that sources exhibiting alternately raising and dimming flux densities may indicate long-period cyclical variability. However, to confirm any hypothesis of this pattern, much longer monitoring over an extended time span would be necessary.

\subsection{Maser flares}

Flares were detected in 10 per cent of sources from our sample (see examples in Figure \ref{Three_sourec_mon}, top and middle and Figure \ref{Appendix_time_series}, top), all which are already identified events in the literature. New short flaring (flickering) events for the G33.641-0.228 59.6~km~s$^{-1}$ line (Figure \ref{Three_sourec_mon}, middle) were detected and continued during our monitoring program, although frequency and amplitudes were typically lower than was reported by \citet{Fujisawa_2014}. We did daily observations during the G24.329+0.144 flare \citep{2019ATel13080....1W} (see Figure \ref{Appendix_time_series}, top).
During the flare, all spectral components exhibited elevated flux levels by factors from 3 to 10. As noted previously, this flare was similar to the one observed 8 years before \citep{Szymczak}, making the G24.329+0.144 an interesting case -- which, to our knowledge, is the first object where a strong, long lasting flare has repeated. This source is of great interest for the Maser Monitoring Organization, and follow--up publications using different data from arrays, like \citet{10.1093/pasj/psac067} and \citet{kobak2023} hopefully will follow.

Most of the periodic and flaring sources in our sample exhibit time lags between different spectral components. This phenomenon is most prominent close to maximum and minimum points. In our sample this is best illustrated in Figures \ref{g107_1flare}. and \ref{Appendix_time_series}, top. 

To identify plausible flares, we suggest the following criteria: the flux measurement should exceed the rolling mean flux (excluding flux measurement's at suspected flare) by at least five standard deviations (\ref{flare_treshold}). Here, $S$ represents the flux, $\overline{S}_{r}$ stands for the rolling mean flux, and $\sigma_{r}$ represents the standard deviation for rolling data. Potential flares can be selected manually or by applying a Z-score test, which involves selecting measurements with scores exceeding 3. Alternatively, other outlier-detection methods may also prove useful. The optimal number of data points to use for calculating the rolling mean and standard deviation can vary depending on the observation intervals and the length of the time series. However, we recommend a minimum of 4 data points, and based on our sample, we have found that using 8 to 14 points demonstrates a high success of identifying maser flares in our sample without giving a false positive for periodic sources.

\begin{equation}
\centering
S > \overline{S}_{r} + 5 \sigma_{r}
\label{flare_treshold}
\end{equation}

To identify the onset of a current flare (one occurring at the present moment), the approach described may not be the most suitable, as it lacks information about the duration and magnitude of the flux increase, and there is no data available for post-flare conditions. In this regard, an empirical approach that defines a flare as a rapid and substantial increase in flux can be more useful. In our opinion, the threshold for identifying a flare can still be determined using the suggested approach by evaluating whether the flux increase still exceeds the recent mean flux density and is five times its standard deviation. 

\subsection{Velocity drift masers}

Only a single source, G121.298+0.659, shows a noticeable velocity drift of the spectral components in the obtained variability time series (Figure \ref{G121_dynamic}.). We observed the spectral component at -25.8~km~s$^{-1}$ drifting towards the -24.9~km~s$^{-1}$ line and merging with it. This process took approximately 500 days, during which the peak flux of the resulting spectral line increased by 30 to 50 percent compared to the initial peak flux density. This may indicate that we observed only the addition of fluxes from individual maser clouds in the velocity domain and not their amplification by positional alignment along the line of sight in the spatial domain (like observed in Orion KL \citealt{Shimoikura_2005} and  G25.65+1.05 \citealt{Burns_2019}). 

Overall, all used variability parameters (variability index, fluctuation index, and $\chi^2_{r}$ parameter) show a noticeable positive correlation. In other words, the most variable sources (with high variability indexes) also exhibit the highest fluctuations and possess high $\chi^2_{r}$ parameters. Source flux density is negatively correlated with variability parameters, meaning that brighter sources tend to be less variable. This result was expected and is in agreement with \citet{Szymczak}. Variability amplitudes and trends in maser sources appear to be equally frequently shared among all spectral components or are equally often unique to individual lines. Typically, prolonged periods of flux decline or increase occur alternately, possibly indicating long source cycles. The findings of our results, including high variability with practically the same spectral components, time delays between light curve extremes, and nearly unnoticeable spectral line velocity changes over five years, strongly support the consensus idea that most of the observed 6.7 GHz methanol maser variability is caused by variations in pumping rates.

\section{Conclusions}
We present the results of a 5-year maser monitoring campaign for 42 sources of 6.7 GHz methanol masers, observed with two radio telescopes located in North West Latvia. Our data on well--known sources show mostly a continuation of previously identified variability trends. Several new variability trends for separate sources were found: possibly periodic fluctuations of G192.60-0.05 and G32.04+0.06; initiation of rapid variability in G78.122+3.633 and G59.783+0.065. Summarising the data also highlighted particular seasonal weather challenges, which are not fully compensated with methods we employed. Due to these limitations, only high amplitude ($\geq30$ per cent) variability can reliably be studied with our telescopes. Our team is committed to enhance the capabilities and accuracy of our facility. These maser monitoring results can be used to select potentially interesting sources for detailed studies using VLBI techniques. 

\section*{Acknowledgements}
We thank the referee for their comments and suggestions which improved the manuscript.
We thank Dr. Juris Kalvans from Ventspils International Radio Astronomy Center, for the valuable comments which improved manuscript.
We thank Dr. Marian Szymczak and his team from Nicolaus Copernicus University, Poland, for the valuable experience transfer observing maser sources with single--dish.
This work has received funding from the ERDF project “Physical and chemical processes in the interstellar medium”, No.1.1.1.1/16/A/213.

\section*{Data Availability}

All source spectra with reasonable request can obtained thought private communication with corresponding author.
Time series plots for all other sources are available in the article and in its online supplementary material.


\bibliographystyle{mnras}
\bibliography{example} 

\begin{thebibliography}{}
\makeatletter
\relax
\def\mn@urlcharsother{\let\do\@makeother \do\$\do\&\do\#\do\^\do\_\do\%\do\~}
\def\mn@doi{\begingroup\mn@urlcharsother \@ifnextchar [ {\mn@doi@}
  {\mn@doi@[]}}
\def\mn@doi@[#1]#2{\def\@tempa{#1}\ifx\@tempa\@empty \href
  {http://dx.doi.org/#2} {doi:#2}\else \href {http://dx.doi.org/#2} {#1}\fi
  \endgroup}
\def\mn@eprint#1#2{\mn@eprint@#1:#2::\@nil}
\def\mn@eprint@arXiv#1{\href {http://arxiv.org/abs/#1} {{\tt arXiv:#1}}}
\def\mn@eprint@dblp#1{\href {http://dblp.uni-trier.de/rec/bibtex/#1.xml}
  {dblp:#1}}
\def\mn@eprint@#1:#2:#3:#4\@nil{\def\@tempa {#1}\def\@tempb {#2}\def\@tempc
  {#3}\ifx \@tempc \@empty \let \@tempc \@tempb \let \@tempb \@tempa \fi \ifx
  \@tempb \@empty \def\@tempb {arXiv}\fi \@ifundefined
  {mn@eprint@\@tempb}{\@tempb:\@tempc}{\expandafter \expandafter \csname
  mn@eprint@\@tempb\endcsname \expandafter{\@tempc}}}

\bibitem[\protect\citeauthoryear{Aberfelds, Shmeld  \& Berzins}{Aberfelds
  et~al.}{2017}]{aberfelds_shmeld_berzins_2017}
Aberfelds A.,  Shmeld I.,   Berzins K.,  2017, \mn@doi [Proceedings of the
  International Astronomical Union] {10.1017/S1743921317009437}, 13, 277

\bibitem[\protect\citeauthoryear{{Aberfelds}, {Steinbergs}, {Shmeld}  \&
  {Bartkiewicz}}{{Aberfelds} et~al.}{2021}]{aberfelds2021}
{Aberfelds} A.,  {Steinbergs} J.,  {Shmeld} I.,   {Bartkiewicz} A.,  2021,
  Astronomical and Astrophysical Transactions, \href
  {https://ui.adsabs.harvard.edu/abs/2021A&AT...32..383A} {32, 383}

\bibitem[\protect\citeauthoryear{Aberfelds, Bartkiewicz, Szymczak, Šteinbergs,
  Surcis, Kobak, Durjasz  \& Shmeld}{Aberfelds et~al.}{2023}]{Aberfelds_2023a}
Aberfelds A.,  Bartkiewicz A.,  Szymczak M.,  Šteinbergs J.,  Surcis G.,
  Kobak A.,  Durjasz M.,   Shmeld I.,  2023, \mn@doi [\mnras]
  {10.1093/mnras/stad1752}, 524, 599

\bibitem[\protect\citeauthoryear{Aller, Aller  \& Hughes}{Aller
  et~al.}{2003}]{Aller_2003}
Aller M.~F.,  Aller H.~D.,   Hughes P.~A.,  2003, \mn@doi [\apj]
  {10.1086/367538}, 586, 33–51

\bibitem[\protect\citeauthoryear{{Antyufeyev} et~al.,}{{Antyufeyev}
  et~al.}{2020}]{2020A&AT...32...23B}
{Antyufeyev} O.,  et~al., 2020, Astronomical and Astrophysical Transactions,
  \href {https://ui.adsabs.harvard.edu/abs/2020A&AT...32...23B} {32, 23}

\bibitem[\protect\citeauthoryear{Araya, Hofner, Goss, Kurtz, Richards, Linz,
  Olmi  \& Sewiło}{Araya et~al.}{2010}]{Araya_2010}
Araya E.~D.,  Hofner P.,  Goss W.~M.,  Kurtz S.,  Richards A. M.~S.,  Linz H.,
  Olmi L.,   Sewiło M.,  2010, \mn@doi [\apjl] {10.1088/2041-8205/717/2/L133},
  717, L133

\bibitem[\protect\citeauthoryear{Bleiders, Bezrukovs  \& Orbidans}{Bleiders
  et~al.}{2017}]{RT-16_performance}
Bleiders M.,  Bezrukovs V.,   Orbidans A.,  2017, \mn@doi [Latvian Journal of
  Physics and Technical Sciences] {doi:10.1515/lpts-2017-0040}, 54, 42

\bibitem[\protect\citeauthoryear{Bleiders, Antyufeyev, Patoka, Orbidans,
  Aberfelds, Steinbergs, Bezrukovs  \& Shmeld}{Bleiders
  et~al.}{2020}]{Bleiders}
Bleiders M.,  Antyufeyev O.,  Patoka O.,  Orbidans A.,  Aberfelds A.,
  Steinbergs J.,  Bezrukovs V.,   Shmeld I.,  2020, \mn@doi [Journal of
  Astronomical Instrumentation] {10.1142/S2251171720500099}, 09, 2050009

\bibitem[\protect\citeauthoryear{Breen et~al.,}{Breen
  et~al.}{2015}]{multibeam_catalog}
Breen S.~L.,  et~al., 2015, \mn@doi [\mnras] {10.1093/mnras/stv847}, 450, 4109

\bibitem[\protect\citeauthoryear{Burns et~al.,}{Burns
  et~al.}{2019}]{Burns_2019}
Burns R.~A.,  et~al., 2019, \mn@doi [Monthly Notices of the Royal Astronomical
  Society] {10.1093/mnras/stz3172}, 491, 4069

\bibitem[\protect\citeauthoryear{{Burns} et~al.,}{{Burns}
  et~al.}{2020}]{burns2020}
{Burns} R.~A.,  et~al., 2020, \mn@doi [Nature Astronomy]
  {10.1038/s41550-019-0989-3}, \href
  {https://ui.adsabs.harvard.edu/abs/2020NatAs...4..506B} {4, 506}

\bibitem[\protect\citeauthoryear{Caratti~o Garatti et~al.,}{Caratti~o Garatti
  et~al.}{2017}]{caratti2017}
Caratti~o Garatti A.,  et~al., 2017, \mn@doi [Nature Physics]
  {10.1038/nphys3942}, \href
  {https://ui.adsabs.harvard.edu/abs/2017NatPh..13..276C} {13, 276}

\bibitem[\protect\citeauthoryear{{Caswell}, {Vaile}  \& {Ellingsen}}{{Caswell}
  et~al.}{1995}]{caswell1995}
{Caswell} J.~L.,  {Vaile} R.~A.,   {Ellingsen} S.~P.,  1995, \mn@doi [\pasa]
  {10.1017/S1323358000020026}, \href
  {https://ui.adsabs.harvard.edu/abs/1995PASA...12...37C} {12, 37}

\bibitem[\protect\citeauthoryear{Durjasz, Szymczak  \& Olech}{Durjasz
  et~al.}{2019}]{durjasz2019}
Durjasz M.,  Szymczak M.,   Olech M.,  2019, \mn@doi [\mnras]
  {10.1093/mnras/stz472}, 485, 777

\bibitem[\protect\citeauthoryear{{Fujisawa} et~al.,}{{Fujisawa}
  et~al.}{2014a}]{fujisawa2014b}
{Fujisawa} K.,  et~al., 2014a, \mn@doi [\pasj] {10.1093/pasj/psu053}, \href
  {https://ui.adsabs.harvard.edu/abs/2014PASJ...66...78F} {66, 78}

\bibitem[\protect\citeauthoryear{Fujisawa et~al.,}{Fujisawa
  et~al.}{2014b}]{Fujisawa_2014}
Fujisawa K.,  et~al., 2014b, \mn@doi [PASJ] {10.1093/pasj/psu097}, 66

\bibitem[\protect\citeauthoryear{{Fujisawa}, {Yonekura}, {Sugiyama},
  {Horiuchi}, {Hayashi}, {Hachisuka}, {Matsumoto}  \& {Niinuma}}{{Fujisawa}
  et~al.}{2015}]{fujisawa2015}
{Fujisawa} K.,  {Yonekura} Y.,  {Sugiyama} K.,  {Horiuchi} H.,  {Hayashi} T.,
  {Hachisuka} K.,  {Matsumoto} N.,   {Niinuma} K.,  2015, The Astronomer's
  Telegram, \href {https://ui.adsabs.harvard.edu/abs/2015ATel.8286....1F}
  {8286, 1}

\bibitem[\protect\citeauthoryear{Goddi, Moscadelli  \& Sanna}{Goddi
  et~al.}{2011}]{Goddi_2011}
Goddi C.,  Moscadelli L.,   Sanna A.,  2011, \mn@doi [A\&A]
  {10.1051/0004-6361/201117854}, 535, L8

\bibitem[\protect\citeauthoryear{{Goedhart}, {Gaylard}  \& {van der
  Walt}}{{Goedhart} et~al.}{2004}]{Goedhart}
{Goedhart} S.,  {Gaylard} M.~J.,   {van der Walt} D.~J.,  2004, \mn@doi
  [\mnras] {10.1111/j.1365-2966.2004.08340.x}, \href
  {https://ui.adsabs.harvard.edu/abs/2004MNRAS.355..553G} {355, 553}

\bibitem[\protect\citeauthoryear{{Goedhart}, {Maswanganye}, {Gaylard}  \& {van
  der Walt}}{{Goedhart} et~al.}{2014}]{goedhart2014b}
{Goedhart} S.,  {Maswanganye} J.~P.,  {Gaylard} M.~J.,   {van der Walt} D.~J.,
  2014, \mn@doi [\mnras] {10.1093/mnras/stt2009}, \href
  {https://ui.adsabs.harvard.edu/abs/2014MNRAS.437.1808G} {437, 1808}

\bibitem[\protect\citeauthoryear{Gray}{Gray}{2012}]{gray_2012}
Gray M.,  2012, Maser Sources in Astrophysics.
Cambridge Astrophysics, Cambridge University Press,
  \mn@doi{10.1017/CBO9780511977534}

\bibitem[\protect\citeauthoryear{{Himwich}}{{Himwich}}{2000}]{2000ivsg.conf...86H}
{Himwich} E.,  2000, in {Takahashi} F.,  ed., International VLBI Service for
  Geodesy and Astrometry 2000 General Meeting Proceedings. pp 86--90

\bibitem[\protect\citeauthoryear{Hirota et~al.,}{Hirota
  et~al.}{2022}]{10.1093/pasj/psac067}
Hirota T.,  et~al., 2022, \mn@doi [PASJ] {10.1093/pasj/psac067}, 74, 1234

\bibitem[\protect\citeauthoryear{Inayoshi, Sugiyama, Hosokawa, Motogi  \&
  Tanaka}{Inayoshi et~al.}{2013}]{Inayoshi_2013}
Inayoshi K.,  Sugiyama K.,  Hosokawa T.,  Motogi K.,   Tanaka K. E.~I.,  2013,
  \mn@doi [\apj] {10.1088/2041-8205/769/2/l20}, 769, L20

\bibitem[\protect\citeauthoryear{{Kobak} et~al.,}{{Kobak}
  et~al.}{2023}]{kobak2023}
{Kobak} A.,  et~al., 2023, \mn@doi [\aap] {10.1051/0004-6361/202244772}, \href
  {https://ui.adsabs.harvard.edu/abs/2023A&A...671A.135K} {671, A135}

\bibitem[\protect\citeauthoryear{{Liu}, {San Liang}  \& {Weisberg}}{{Liu}
  et~al.}{2007}]{Liu_2007}
{Liu} Y.,  {San Liang} X.,   {Weisberg} R.~H.,  2007, \mn@doi [Journal of
  Atmospheric and Oceanic Technology] {10.1175/2007JTECHO511.1}, \href
  {https://ui.adsabs.harvard.edu/abs/2007JAtOT..24.2093L} {24, 2093}

\bibitem[\protect\citeauthoryear{{MacLeod} et~al.,}{{MacLeod}
  et~al.}{2018}]{macleod2018}
{MacLeod} G.~C.,  et~al., 2018, \mn@doi [\mnras] {10.1093/mnras/sty996}, \href
  {https://ui.adsabs.harvard.edu/abs/2018MNRAS.478.1077M} {478, 1077}

\bibitem[\protect\citeauthoryear{{Menten}}{{Menten}}{1991}]{menten1991}
{Menten} K.~M.,  1991, \mn@doi [\apjl] {10.1086/186177}, \href
  {https://ui.adsabs.harvard.edu/abs/1991ApJ...380L..75M} {380, L75}

\bibitem[\protect\citeauthoryear{Moscadelli et~al.,}{Moscadelli
  et~al.}{2017}]{Moscadelli_2017}
Moscadelli L.,  et~al., 2017, \mn@doi [A\&A] {10.1051/0004-6361/201730659},
  600, L8

\bibitem[\protect\citeauthoryear{{Olech}, {Szymczak}, {Wolak}, {Sarniak}  \&
  {Bartkiewicz}}{{Olech} et~al.}{2019}]{olech2019}
{Olech} M.,  {Szymczak} M.,  {Wolak} P.,  {Sarniak} R.,   {Bartkiewicz} A.,
  2019, \mn@doi [\mnras] {10.1093/mnras/stz926}, \href
  {https://ui.adsabs.harvard.edu/abs/2019MNRAS.486.1236O} {486, 1236}

\bibitem[\protect\citeauthoryear{{Olech}, {Szymczak}, {Wolak}, {G{\'e}rard}  \&
  {Bartkiewicz}}{{Olech} et~al.}{2020}]{olech2020}
{Olech} M.,  {Szymczak} M.,  {Wolak} P.,  {G{\'e}rard} E.,   {Bartkiewicz} A.,
  2020, \mn@doi [\aap] {10.1051/0004-6361/201936943}, \href
  {https://ui.adsabs.harvard.edu/abs/2020A&A...634A..41O} {634, A41}

\bibitem[\protect\citeauthoryear{{Olech}, {Durjasz}, {Szymczak}  \&
  {Bartkiewicz}}{{Olech} et~al.}{2022}]{olech2022}
{Olech} M.,  {Durjasz} M.,  {Szymczak} M.,   {Bartkiewicz} A.,  2022, \mn@doi
  [\aap] {10.1051/0004-6361/202243108}, \href
  {https://ui.adsabs.harvard.edu/abs/2022A&A...661A.114O} {661, A114}

\bibitem[\protect\citeauthoryear{Parfenov \& Sobolev}{Parfenov \&
  Sobolev}{2014}]{Parfenov_2014}
Parfenov S.~Y.,  Sobolev A.~M.,  2014, \mn@doi [\mnras]
  {10.1093/mnras/stu1481}, 444, 620

\bibitem[\protect\citeauthoryear{Perley \& Butler}{Perley \&
  Butler}{2013}]{Perley_2013}
Perley R.~A.,  Butler B.~J.,  2013, \mn@doi [A\&AS]
  {10.1088/0067-0049/204/2/19}, 204, 19

\bibitem[\protect\citeauthoryear{Ryabov \& Vrublevskis}{Ryabov \&
  Vrublevskis}{2023}]{Ryabov_2023}
Ryabov B.,  Vrublevskis A.,  2023, \mn@doi [Latvian Journal of Physics and
  Technical Sciences] {doi:10.2478/lpts-2023-0011}, 60, 52

\bibitem[\protect\citeauthoryear{Salii, Zinchenko, Liu, Sobolev, Aberfelds  \&
  Su}{Salii et~al.}{2022}]{salii_2022}
Salii S.~V.,  Zinchenko I.~I.,  Liu S.-Y.,  Sobolev A.~M.,  Aberfelds A.,   Su
  Y.-N.,  2022, \mn@doi [\mnras] {10.1093/mnras/stac739}, 512, 3215

\bibitem[\protect\citeauthoryear{{Scargle}}{{Scargle}}{1982}]{scargle1982}
{Scargle} J.~D.,  1982, \mn@doi [\apj] {10.1086/160554}, \href
  {http://adsabs.harvard.edu/abs/1982ApJ...263..835S} {263, 835}

\bibitem[\protect\citeauthoryear{{Sebastian}, {Nabil}, {Alexey}, {Christopher}
  \& {Gilbert}}{{Sebastian} et~al.}{2023}]{Sebastian2023}
{Sebastian} K.,  {Nabil} F.,  {Alexey} B.,  {Christopher} T.,   {Gilbert}
  P.~C.,  2023, PyCWT, \url{https://github.com/regeirk/pycwt}

\bibitem[\protect\citeauthoryear{{Shimoikura}, {Kobayashi}, {Omodaka},
  {Diamond}, {Matveyenko}  \& {Fujisawa}}{{Shimoikura}
  et~al.}{2005}]{Shimoikura_2005}
{Shimoikura} T.,  {Kobayashi} H.,  {Omodaka} T.,  {Diamond} P.~J.,
  {Matveyenko} L.~I.,   {Fujisawa} K.,  2005, \mn@doi [The Astrophysical
  Journal] {10.1086/432865}, \href
  {https://ui.adsabs.harvard.edu/abs/2005ApJ...634..459S} {634, 459}

\bibitem[\protect\citeauthoryear{{Stecklum} et~al.,}{{Stecklum}
  et~al.}{2021}]{stecklum2021}
{Stecklum} B.,  et~al., 2021, \mn@doi [\aap] {10.1051/0004-6361/202039645},
  \href {https://ui.adsabs.harvard.edu/abs/2021A&A...646A.161S} {646, A161}

\bibitem[\protect\citeauthoryear{{Steinbergs}, {Aberfelds}, {Bleiders}  \&
  {Shmeld}}{{Steinbergs} et~al.}{2021}]{Maser-Data-Processing-Suite}
{Steinbergs} J.,  {Aberfelds} A.,  {Bleiders} M.,   {Shmeld} I.,  2021, VIRAC
  maser data processing suite, \url {https://doi.org/10.17184/eac.5643}

\bibitem[\protect\citeauthoryear{{Stetson}}{{Stetson}}{1996}]{Stetson_1996}
{Stetson} P.~B.,  1996, \mn@doi [\pasp] {10.1086/133808}, \href
  {https://ui.adsabs.harvard.edu/abs/1996PASP..108..851S} {108, 851}

\bibitem[\protect\citeauthoryear{Sugiyama et~al.,}{Sugiyama
  et~al.}{2019a}]{Sugiyama2019}
Sugiyama K.,  et~al., 2019a, \mn@doi [Journal of Physics: Conference Series]
  {10.1088/1742-6596/1380/1/012057}, 1380, 012057

\bibitem[\protect\citeauthoryear{Sugiyama, Saito, Yonekura  \& Momose}{Sugiyama
  et~al.}{2019b}]{Sugiyama_2019}
Sugiyama K.,  Saito Y.,  Yonekura Y.,   Momose M.,  2019b, The Astron. Telegr.,
  12446, 1

\bibitem[\protect\citeauthoryear{{Sukharev}, {Ryabov}, {Bezrukovs}  \&
  {Orbidans}}{{Sukharev} et~al.}{2022}]{Sukharev_2022}
{Sukharev} A.,  {Ryabov} M.,  {Bezrukovs} V.,   {Orbidans} A.,  2022,
  Astronomical and Astrophysical Transactions, \href
  {https://ui.adsabs.harvard.edu/abs/2022A&AT...33..149S} {33, 149}

\bibitem[\protect\citeauthoryear{Szymczak, Wolak, Bartkiewicz  \& van
  Langevelde}{Szymczak et~al.}{2011}]{Szymczak_2011}
Szymczak M.,  Wolak P.,  Bartkiewicz A.,   van Langevelde H.~J.,  2011, \mn@doi
  [A\&A] {10.1051/0004-6361/201117145}, 531, L3

\bibitem[\protect\citeauthoryear{{Szymczak}, {Wolak}, {Bartkiewicz}  \&
  {Borkowski}}{{Szymczak} et~al.}{2012}]{Toruncatalog}
{Szymczak} M.,  {Wolak} P.,  {Bartkiewicz} A.,   {Borkowski} K.~M.,  2012,
  \mn@doi [Astronomische Nachrichten] {10.1002/asna.201211702}, \href
  {https://ui.adsabs.harvard.edu/abs/2012AN....333..634S} {333, 634}

\bibitem[\protect\citeauthoryear{{Szymczak}, {Wolak}  \&
  {Bartkiewicz}}{{Szymczak} et~al.}{2014}]{szymczak2014}
{Szymczak} M.,  {Wolak} P.,   {Bartkiewicz} A.,  2014, \mn@doi [\mnras]
  {10.1093/mnras/stu019}, \href
  {https://ui.adsabs.harvard.edu/abs/2014MNRAS.439..407S} {439, 407}

\bibitem[\protect\citeauthoryear{Szymczak, Wolak  \& Bartkiewicz}{Szymczak
  et~al.}{2015}]{Szymczak_2015}
Szymczak M.,  Wolak P.,   Bartkiewicz A.,  2015, \mn@doi [\mnras]
  {10.1093/mnras/stv145}, \href
  {https://ui.adsabs.harvard.edu/abs/2015MNRAS.448.2284S} {448, 2284}

\bibitem[\protect\citeauthoryear{Szymczak, Olech, Wolak, Bartkiewicz  \&
  Gawroński}{Szymczak et~al.}{2016}]{Szymczak_2016}
Szymczak M.,  Olech M.,  Wolak P.,  Bartkiewicz A.,   Gawroński M.,  2016,
  \mn@doi [\mnras] {10.1093/mnrasl/slw044}, 459, L56–L60

\bibitem[\protect\citeauthoryear{{Szymczak}, {Olech}, {Sarniak}, {Wolak}  \&
  {Bartkiewicz}}{{Szymczak} et~al.}{2018}]{Szymczak}
{Szymczak} M.,  {Olech} M.,  {Sarniak} R.,  {Wolak} P.,   {Bartkiewicz} A.,
  2018, \mn@doi [\mnras] {10.1093/mnras/stx2693}, \href
  {https://ui.adsabs.harvard.edu/abs/2018MNRAS.474..219S} {474, 219}

\bibitem[\protect\citeauthoryear{{Torrence} \& {Compo}}{{Torrence} \&
  {Compo}}{1998}]{Torrence_1998}
{Torrence} C.,  {Compo} G.~P.,  1998, \mn@doi [Bulletin of the American
  Meteorological Society] {10.1175/1520-0477(1998)079<0061:APGTWA>2.0.CO;2},
  \href {https://ui.adsabs.harvard.edu/abs/1998BAMS...79...61T} {79, 61}

\bibitem[\protect\citeauthoryear{Virtanen et~al.,}{Virtanen
  et~al.}{2020}]{2020SciPy-NMeth}
Virtanen P.,  et~al., 2020, \mn@doi [Nature Methods]
  {10.1038/s41592-019-0686-2}, \href {https://rdcu.be/b08Wh} {17, 261}

\bibitem[\protect\citeauthoryear{{Walsh}, {Burton}, {Hyland}  \&
  {Robinson}}{{Walsh} et~al.}{1998}]{Walsh}
{Walsh} A.~J.,  {Burton} M.~G.,  {Hyland} A.~R.,   {Robinson} G.,  1998,
  \mn@doi [\mnras] {10.1046/j.1365-8711.1998.02014.x}, \href
  {https://ui.adsabs.harvard.edu/abs/1998MNRAS.301..640W} {301, 640}

\bibitem[\protect\citeauthoryear{Winkel, Kraus  \& Bach}{Winkel
  et~al.}{2012}]{Winkel}
Winkel B.,  Kraus A.,   Bach U.,  2012, \mn@doi [A\&A]
  {10.1051/0004-6361/201118092}, 540, A140

\bibitem[\protect\citeauthoryear{{Wolak}, {Olech}, {Szymczak}, {Bartkiewicz}
  \& {Durjasz}}{{Wolak} et~al.}{2019}]{2019ATel13080....1W}
{Wolak} P.,  {Olech} M.,  {Szymczak} M.,  {Bartkiewicz} A.,   {Durjasz} M.,
  2019, The Astronomer's Telegram, \href
  {https://ui.adsabs.harvard.edu/abs/2019ATel13080....1W} {13080, 1}

\bibitem[\protect\citeauthoryear{Yonekura et~al.,}{Yonekura
  et~al.}{2016}]{Yonekura2016}
Yonekura Y.,  et~al., 2016, \mn@doi [Publications of the Astronomical Society
  of Japan] {10.1093/pasj/psw045}, 68, 74

\bibitem[\protect\citeauthoryear{Younes et~al.,}{Younes
  et~al.}{2022}]{Younes_2022}
Younes G.,  et~al., 2022, \mn@doi [\apj] {10.3847/1538-4357/ac3756}, 924, 136

\bibitem[\protect\citeauthoryear{{Zinnecker} \& {Yorke}}{{Zinnecker} \&
  {Yorke}}{2007}]{Zinnecker}
{Zinnecker} H.,  {Yorke} H.~W.,  2007, \mn@doi [\araa]
  {10.1146/annurev.astro.44.051905.092549}, \href
  {https://ui.adsabs.harvard.edu/abs/2007ARA&A..45..481Z} {45, 481}

\bibitem[\protect\citeauthoryear{van~der Walt}{van~der
  Walt}{2011}]{van_der_Walt_2011}
van~der Walt D.~J.,  2011, \mn@doi [\apj] {10.1088/0004-6256/141/5/152}, 141,
  152

\bibitem[\protect\citeauthoryear{van~der Walt, Sobolev  \& Butner}{van~der Walt
  et~al.}{2007}]{Walt}
van~der Walt D.~J.,  Sobolev A.~M.,   Butner H.,  2007, \mn@doi [A\&A]
  {10.1051/0004-6361:20065638}, 464, 1015

\makeatother
\end{thebibliography}




\appendix

\section{Additional time series}

\begin{figure*}
\centering
\includegraphics[width=0.85\paperwidth, height=0.22\textheight]{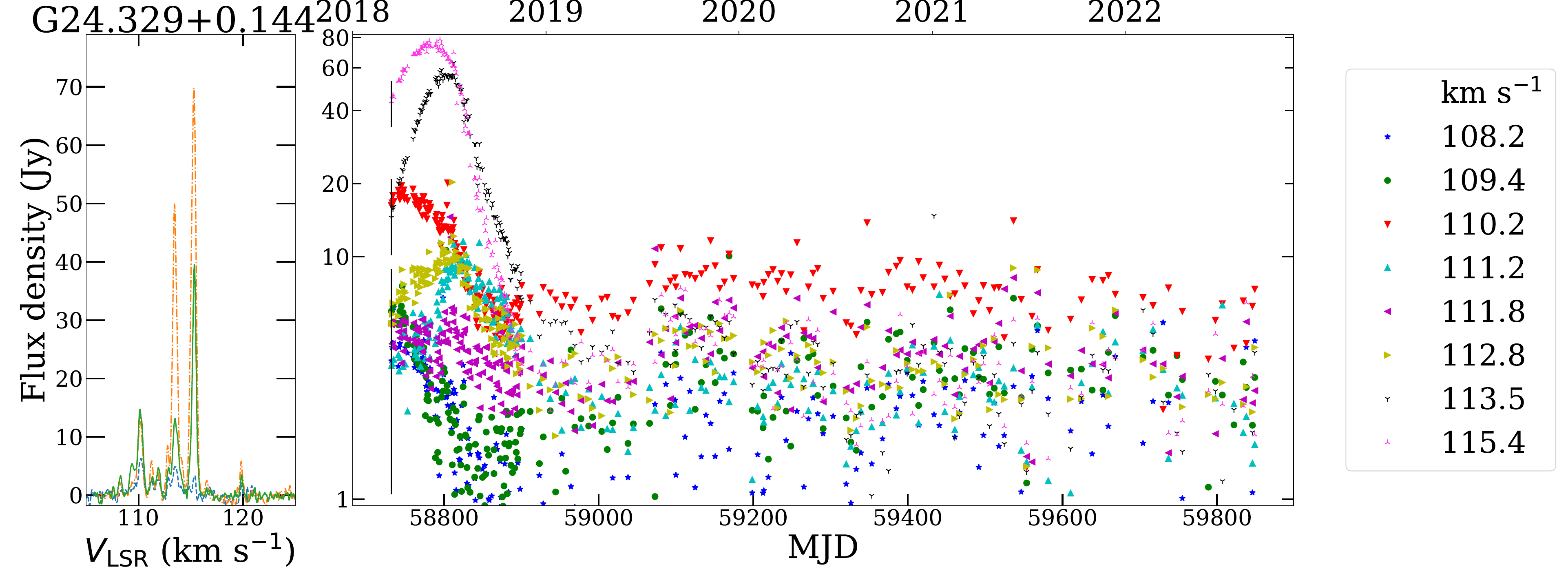}
\includegraphics[width=0.85\paperwidth, height=0.22\textheight]{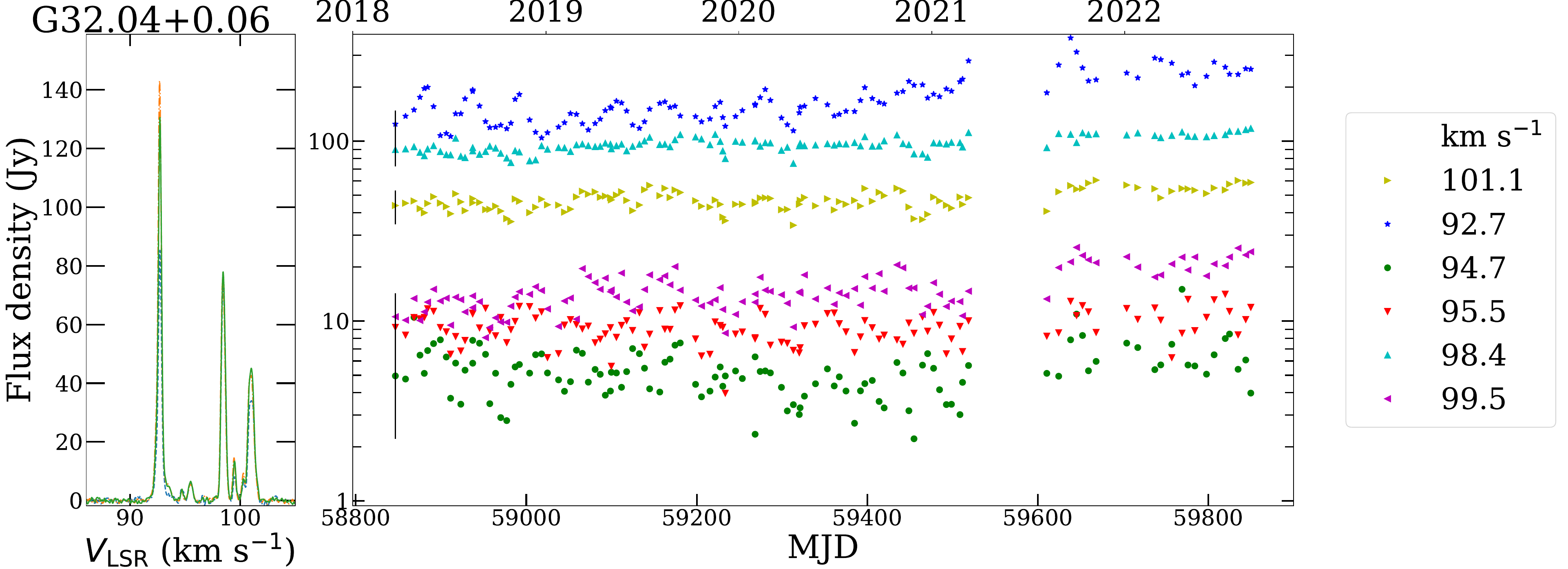}
\includegraphics[width=0.85\paperwidth, height=0.22\textheight]{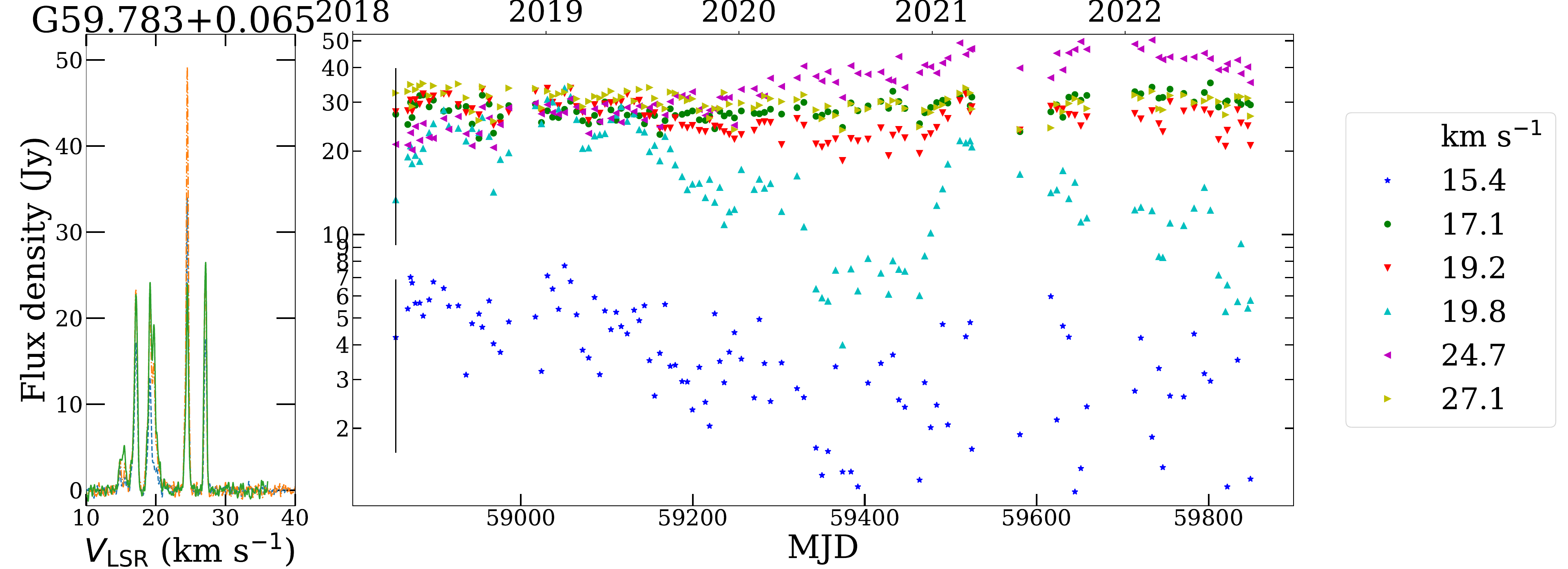}
\caption{Same as Figure \ref{Three_sourec_mon}., but for the G24.329+0.144; G32.04+0.06 and G59.783+0.065. } 
\label{Appendix_time_series}
\end{figure*}

\begin{figure*}
\centering
\includegraphics[width=0.8\paperwidth]{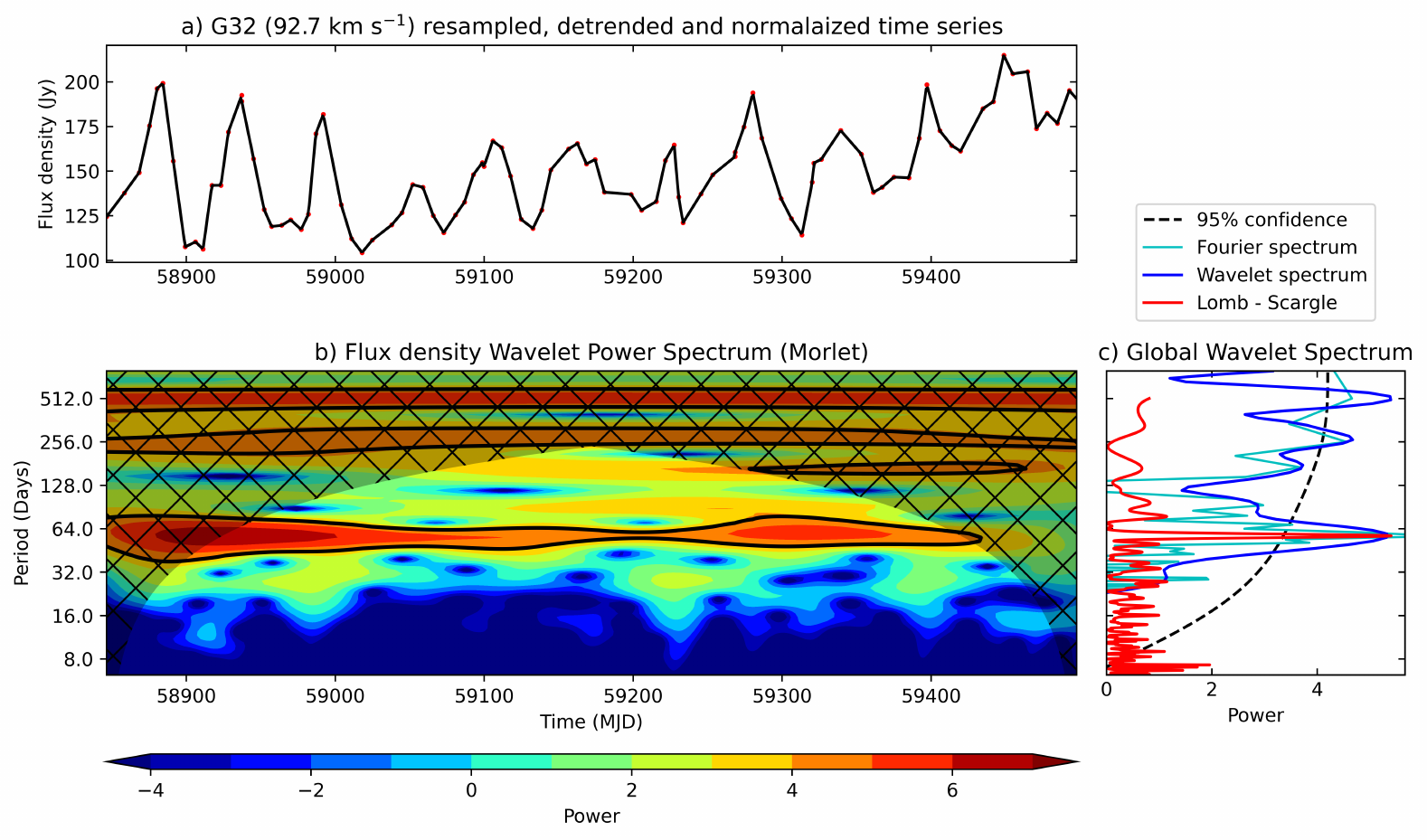}
\caption{Same as Figure \ref{g107_7.4_wavelet}., but for the G32.04+0.06 92.7~km~s$^{-1}$ line. Note, only data obtained until 59500 MJD (October 2021) were used.  } 
\label{G32p04_wavelet}
\end{figure*}

\begin{figure*}
\centering
\includegraphics[scale=0.2]{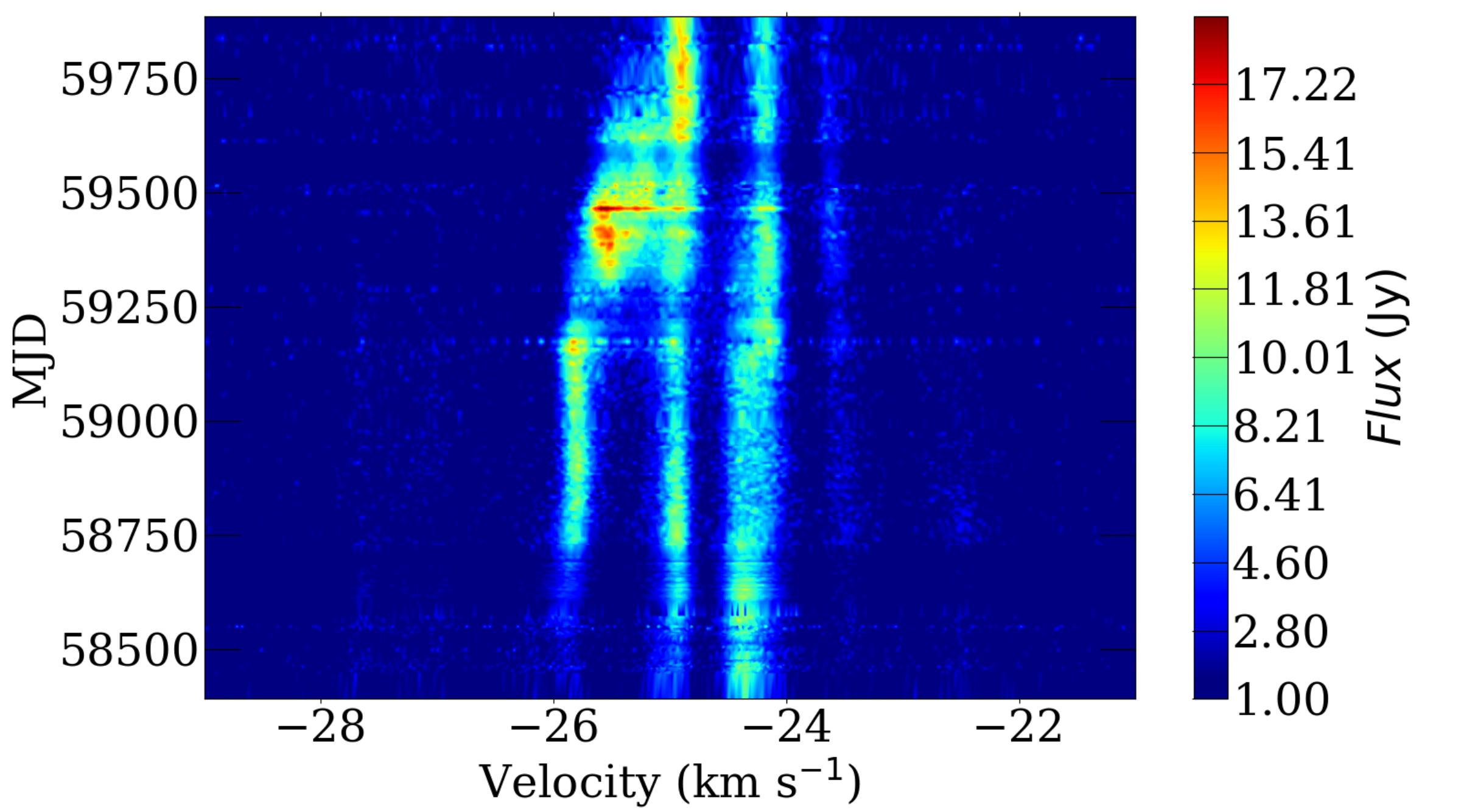}
\caption{Dynamic spectra of G121.298+0.659 illustrating velocity drifting. } 
\label{G121_dynamic}
\end{figure*}
\pagebreak
\newpage
\section{Full variability parameter table}

\label{full_var_table}
\tablecaption{Full table of maser line variability parameters, same information as in Table \ref{variability parametrs}}
\tablefirsthead{\toprule V\textsubscript{P}(km s$^{-1}$) & S\textsubscript{P}(Jy) & $VI$ & $FI$ & $\chi^2_{r}$ \\ \midrule}
\tablehead{%
\multicolumn{5}{c}%
{{\bfseries  Continued from previous column}} \\
\toprule
V\textsubscript{P}(km s$^{-1}$) & S\textsubscript{P}(Jy) & $VI$ & $FI$ & $\chi^2_{r}$\\ \midrule}
\tabletail{%
\midrule \multicolumn{5}{r}{{Continued on next column}} \\ \midrule} 
\tablelasttail{
\multicolumn{5}{c}{{Concluded}}}
\begin{xtabular}{l c c c l}
\multicolumn{5}{l}{G22.357+0.066 (MJD\textsubscript{s}=58220, $T\textsubscript{s}$= 4.462\,y, $N$=202)} \\
78.7 &  2.60 & 0.74 & 0.70 & 0.84\\
80.1 &  18.49 & 0.40 & 0.46 & 1.11\\
88.3 &  2.83 & 0.58 & 0.53 & 0.59\\
\hline
\multicolumn{5}{l}{G24.329+0.144 (MJD\textsubscript{s}=58731, $T\textsubscript{s}$= 3.060\,y, $N$=236)} \\
108.2 &  2.12 & 0.19 & 0.64 & 0.37\\
109.4 &  2.82 & 0.47 & 0.57 & 0.42\\
110.2 &  9.30 & 0.73 & 0.49 & 1.96\\
111.2 &  4.70 & 0.61 & 0.66 & 1.05\\
111.8 &  4.13 & 0.76 & 0.41 & 0.36\\
112.8 &  5.58 & 0.80 & 0.62 & 1.22\\
113.5 &  18.73 & 0.94 & 1.50 & 33.70\\
115.4 &  23.76 & 0.88 & 1.90 & 58.40\\
\hline
\multicolumn{5}{l}{G25.709+0.044 (MJD\textsubscript{s}=58846, $T\textsubscript{s}$= 2.751\,y, $N$=108)} \\
89.8 &  50.14 & 0.04 & 0.09 & 0.15\\
91.3 &  12.65 & 0.16 & 0.22 & 0.27\\
92.0 &  26.63 & 0.07 & 0.14 & 0.20\\
93.3 &  43.55 & -0.05 & 0.17 & 0.28\\
94.0 &  64.57 & -0.03 & 0.12 & 0.24\\
95.5 &  321.65 & -0.04 & 0.06 & 0.22\\
95.7 &  328.85 & -0.01 & 0.07 & 0.17\\
97.9 &  7.48 & 0.12 & 0.28 & 0.15\\
\hline
\multicolumn{5}{l}{G25.64+1.05 (MJD\textsubscript{s}=58024, $T\textsubscript{s}$= 5.003\,y, $N$=253)} \\
38.6 &  2.42 & 0.43 & 0.89 & 0.34\\
40.3 &  4.38 & 0.15 & 0.56 & 0.34\\
41.8 &  129.75 & 0.07 & 0.17 & 0.99\\
42.5 &  3.06 & 0.35 & 0.58 & 0.36\\
\hline
\multicolumn{5}{l}{G30.99-0.08 (MJD\textsubscript{s}=57950, $T\textsubscript{s}$= 5.202\,y, $N$=248)} \\
74.8 &  7.91 & 0.01 & 0.36 & 0.81\\
77.8 &  12.69 & 0.45 & 0.59 & 3.48\\
88.4 &  4.87 & 0.63 & 0.61 & 1.55\\
\hline
\multicolumn{5}{l}{G32.04+0.06 (MJD\textsubscript{s}=58846, $T\textsubscript{s}$= 2.748\,y, $N$=117)} \\
101.1 &  47.37 & 0.10 & 0.16 & 0.35\\
92.7 &  172.25 & 0.40 & 0.50 & 2.50\\
94.7 &  5.36 & 0.47 & 0.51 & 0.45\\
95.5 &  9.30 & 0.23 & 0.28 & 0.39\\
98.4 &  96.06 & 0.07 & 0.07 & 0.28\\
99.5 &  15.12 & 0.25 & 0.38 & 0.76\\
\hline
\multicolumn{5}{l}{G32.744-0.076 (MJD\textsubscript{s}=57825, $T\textsubscript{s}$= 5.596\,y, $N$=1016)} \\
30.5 &  15.01 & -0.04 & 0.14 & 0.15\\
32.1 &  30.67 & 0.15 & 0.06 & 0.43\\
33.5 &  38.56 & 0.01 & 0.08 & 0.23\\
38.1 &  37.89 & 0.09 & 0.11 & 0.26\\
38.3 &  40.43 & 0.10 & 0.04 & 0.21\\
38.5 &  42.97 & 0.03 & 0.07 & 0.26\\
39.2 &  16.88 & 0.09 & 0.07 & 0.17\\
\hline
\multicolumn{5}{l}{G33.641-0.228 (MJD\textsubscript{s}=57821, $T\textsubscript{s}$= 5.707\,y, $N$=1136)} \\
59.3 &  18.78 & 0.83 & 1.05 & 17.86\\
59.6 &  15.40 & 0.73 & 0.77 & 2.18\\
60.3 &  161.73 & 0.21 & 0.50 & 2.66\\
61.0 &  49.98 & 0.49 & 0.65 & 4.56\\
62.7 &  18.39 & 0.32 & 0.18 & 0.56\\
63.2 &  15.21 & 0.39 & 0.16 & 0.28\\
\hline
\multicolumn{5}{l}{G35.20-1.74 (MJD\textsubscript{s}=57956, $T\textsubscript{s}$= 5.186\,y, $N$=277)} \\
41.0 &  3.11 & 0.81 & 0.58 & 1.50\\
43.1 &  23.70 & 0.19 & 0.30 & 0.66\\
43.8 &  14.15 & 0.62 & 0.45 & 5.42\\
44.1 &  11.79 & 0.52 & 0.34 & 2.08\\
44.6 &  35.31 & 0.08 & 0.18 & 0.33\\
44.9 &  25.62 & 0.14 & 0.18 & 0.73\\
45.3 &  17.51 & 0.02 & 0.26 & 0.54\\
\hline
\multicolumn{5}{l}{G34.396+0.222 (MJD\textsubscript{s}=58312, $T\textsubscript{s}$= 4.210\,y, $N$=204)} \\
55.6 &  24.10 & 0.42 & 0.52 & 2.34\\
62.5 &  5.55 & 0.17 & 0.27 & 0.25\\
\hline
\multicolumn{5}{l}{G36.705+0.096 (MJD\textsubscript{s}=58289, $T\textsubscript{s}$= 4.273\,y, $N$=191)} \\
62.2 &  2.56 & 0.25 & 0.43 & 0.41\\
\hline
\multicolumn{5}{l}{G37.479-0.105 (MJD\textsubscript{s}=57881, $T\textsubscript{s}$= 5.393\,y, $N$=256)} \\
50.0 &  2.58 & -0.31 & 0.17 & 0.55\\
52.3 &  1.87 & -0.51 & 0.54 & 0.41\\
56.9 &  9.47 & -0.06 & 0.18 & 0.32\\
58.3 &  4.85 & -0.07 & 0.05 & 0.47\\
59.3 &  4.90 & -0.16 & 0.25 & 0.24\\
60.1 &  1.99 & 0.21 & 0.46 & 0.29\\
62.0 &  9.11 & 0.13 & 0.35 & 0.28\\
62.5 &  7.38 & -0.06 & 0.38 & 0.26\\
\hline
\multicolumn{5}{l}{G37.43+1.52 (MJD\textsubscript{s}=58846, $T\textsubscript{s}$= 2.748\,y, $N$=111)} \\
41.3 &  360.69 & 0.03 & 0.10 & 0.21\\
\hline
\multicolumn{5}{l}{G37.55+0.20 (MJD\textsubscript{s}=58298, $T\textsubscript{s}$= 4.246\,y, $N$=198)} \\
83.8 &  6.25 & 0.40 & 0.56 & 1.13\\
84.8 &  5.81 & 0.20 & 0.42 & 0.73\\
86.4 &  3.60 & 0.28 & 0.34 & 0.51\\
\hline
\multicolumn{5}{l}{G43.149+0.013 (MJD\textsubscript{s}=57881, $T\textsubscript{s}$= 5.398\,y, $N$=265)} \\
13.2 &  16.11 & -0.04 & 0.15 & 0.18\\
18.9 &  20.51 & -0.06 & 0.14 & 0.30\\
19.6 &  12.24 & 0.16 & 0.32 & 0.33\\
20.2 &  8.53 & 0.03 & 0.23 & 0.28\\
8.4 &  16.17 & -0.02 & 0.22 & 0.23\\
9.3 &  37.00 & 0.03 & 0.19 & 0.29\\
\hline
\multicolumn{5}{l}{G43.796-0.127 (MJD\textsubscript{s}=58319, $T\textsubscript{s}$= 4.189\,y, $N$=204)} \\
39.6 &  15.97 & 0.18 & 0.16 & 0.41\\
40.0 &  16.71 & 0.14 & 0.15 & 0.47\\
40.4 &  19.48 & 0.00 & 0.08 & 0.30\\
43.0 &  15.27 & 0.03 & 0.15 & 0.23\\
\hline
\multicolumn{4}{l}{G45.071+0.013 (MJD\textsubscript{s}=58854, $T\textsubscript{s}$= 2.724\,y, $N$=112)} \\
57.8 &  40.06 & -0.12 & 0.10 & 0.16\\
\hline
\multicolumn{5}{l}{G49.04-1.08 (MJD\textsubscript{s}=58389, $T\textsubscript{s}$= 1.251\,y, $N$=61)} \\
35.6 &  14.25 & 0.14 & 0.20 & 0.95\\
36.5 &  8.49 & 0.24 & 0.13 & 0.46\\
37.1 &  13.47 & 0.46 & 0.34 & 2.96\\
38.4 &  4.53 & 0.65 & 1.12 & 5.25\\
39.2 &  5.99 & 0.57 & 0.65 & 3.96\\
40.9 &  1.43 & 0.53 & 0.67 & 1.10\\
\hline
\multicolumn{5}{l}{G196.454-01.67 (MJD\textsubscript{s}=58842, $T\textsubscript{s}$= 2.774\,y, $N$=88)} \\
14.7 &  13.85 & 0.60 & 0.76 & 2.85\\
15.2 &  10.64 & 0.35 & 0.42 & 0.89\\
15.5 &  10.26 & 0.52 & 0.59 & 1.56\\
\hline
\multicolumn{5}{l}{G49.490-0.388 (MJD\textsubscript{s}=57885, $T\textsubscript{s}$= 5.388,}\\
\multicolumn{5}{l}{$N$=297, $C(month^{-1})$=4.594)} \\
50.1 &  11.87 & 0.19 & 0.23 & 0.21\\
51.8 &  55.05 & 0.03 & 0.22 & 0.34\\
56.2 &  39.85 & 0.04 & 0.25 & 0.44\\
57.9 &  128.99 & 0.22 & 0.22 & 1.61\\
58.3 &  147.05 & 0.02 & 0.23 & 0.83\\
58.8 &  274.17 & 0.41 & 0.49 & 5.37\\
59.3 &  842.08 & 0.00 & 0.14 & 0.49\\
\hline
\multicolumn{5}{l}{G192.60-0.05 (MJD\textsubscript{s}=57856, $T\textsubscript{s}$= 5.476\,y, $N$=368)} \\
2.3 &  7.28 & 0.92 & 2.52 & 8.39\\
4.2 &  7.17 & 0.67 & 0.79 & 1.71\\
4.8 &  15.74 & 0.63 & 0.75 & 2.56\\
5.9 &  176.49 & 0.37 & 0.52 & 3.05\\
6.3 &  162.01 & 0.81 & 1.59 & 17.18\\
7.5 &  4.51 & 0.97 & 4.24 & 6.32\\
\hline
\multicolumn{5}{l}{G189.030+0.784 (MJD\textsubscript{s}=57839, $T\textsubscript{s}$= 2.567\,y, $N$=155)} \\
8.8 &  16.80 & 0.55 & 0.12 & 1.03\\
9.6 &  14.55 & 0.28 & 0.28 & 2.48\\
\hline
\multicolumn{5}{l}{G59.783+0.065 (MJD\textsubscript{s}=58854, $T\textsubscript{s}$= 2.724\,y, $N$=113)} \\
15.4 &  3.62 & 0.56 & 0.60 & 0.98\\
17.1 &  28.46 & -0.05 & 0.12 & 0.14\\
19.2 &  26.52 & -0.02 & 0.26 & 0.31\\
19.8 &  16.44 & 0.59 & 0.79 & 4.60\\
24.7 &  33.57 & 0.21 & 0.38 & 1.17\\
27.1 &  30.28 & -0.08 & 0.15 & 0.13\\
\hline
\multicolumn{5}{l}{G69.540-0.976 (MJD\textsubscript{s}=57832, $T\textsubscript{s}$= 5.530\,y, $N$=206)} \\
0.0 &  6.54 & 0.42 & 0.19 & 0.47\\
1.3 &  3.95 & 0.16 & 0.51 & 0.35\\
14.6 &  112.12 & 0.15 & 0.16 & 0.44\\
\hline
\multicolumn{5}{l}{G174.20-0.08 (MJD\textsubscript{s}=57832, $T\textsubscript{s}$= 5.542\,y, $N$=261)} \\
1.4 &  43.69 & 0.50 & 0.37 & 3.88\\
3.8 &  9.27 & 0.60 & 0.56 & 3.30\\
4.6 &  5.19 & 0.15 & 0.24 & 0.40\\
\hline
\multicolumn{5}{l}{G173.482+2.446 (MJD\textsubscript{s}=57821, $T\textsubscript{s}$= 5.572\,y, $N$=250)} \\
-11.8 &  6.78 & 0.18 & 0.40 & 0.49\\
-13.0 &  25.46 & 0.36 & 0.29 & 1.57\\
-13.8 &  18.91 & 0.15 & 0.43 & 1.02\\
-7.5 &  9.86 & 0.41 & 0.73 & 3.78\\
\hline
\multicolumn{5}{l}{G73.06+1.80 (MJD\textsubscript{s}=58859, $T\textsubscript{s}$= 2.709\,y, $N$=137)} \\
-2.7 &  1.65 & 0.68 & 0.76 & 0.67\\
6.0 &  7.42 & 0.36 & 0.55 & 1.20\\
\hline
\multicolumn{5}{l}{G75.782+0.343 (MJD\textsubscript{s}=57885, $T\textsubscript{s}$= 5.379\,y, $N$=313)} \\
-0.1 &  4.86 & 0.68 & 0.58 & 2.18\\
-0.4 &  6.68 & 0.35 & 0.28 & 0.92\\
-2.6 &  63.86 & 0.14 & 0.04 & 0.69\\
0.7 &  25.71 & 0.42 & 0.22 & 2.35\\
\hline
\multicolumn{5}{l}{G78.122+3.633 (MJD\textsubscript{s}=57832, $T\textsubscript{s}$= 5.549\,y, $N$=882)} \\
-6.1 &  27.70 & 0.24 & 0.20 & 0.61\\
-6.7 &  29.15 & 0.74 & 0.67 & 11.66\\
-7.0 &  18.42 & 0.62 & 0.64 & 4.87\\
-7.7 &  24.06 & 0.97 & 2.48 & 65.51\\
\hline
\multicolumn{5}{l}{G81.88+0.78 (MJD\textsubscript{s}=57834, $T\textsubscript{s}$= 5.525\,y, $N$=300)} \\
3.5 &  83.53 & 0.33 & 0.42 & 1.34\\
4.1 &  219.65 & 0.38 & 0.47 & 1.92\\
4.6 &  311.00 & 0.16 & 0.27 & 0.80\\
5.2 &  105.18 & 0.17 & 0.22 & 0.57\\
5.8 &  99.84 & 0.15 & 0.24 & 0.62\\
7.3 &  264.17 & 0.34 & 0.37 & 1.92\\
9.5 &  36.10 & 0.52 & 0.66 & 2.87\\
\hline
\multicolumn{5}{l}{G188.95+0.89 (MJD\textsubscript{s}=57993, $T\textsubscript{s}$= 5.101\,y, $N$=223)} \\
10.5 &  478.64 & 0.28 & 0.23 & 1.60\\
10.8 &  868.89 & 0.11 & 0.32 & 1.30\\
11.6 &  13.51 & 0.46 & 0.37 & 1.18\\
11.8 &  8.32 & 0.71 & 0.52 & 2.88\\
8.5 &  6.88 & 0.17 & 0.28 & 0.36\\
9.7 &  39.52 & 0.06 & 0.29 & 0.75\\
\hline
\multicolumn{5}{l}{G85.411+0.002 (MJD\textsubscript{s}=58842, $T\textsubscript{s}$= 2.832\,y, $N$=698)} \\
-28.6 &  5.15 & 0.34 & 0.38 & 0.38\\
-29.4 &  84.99 & 0.01 & 0.16 & 0.21\\
-30.8 &  3.98 & 0.36 & 0.37 & 0.32\\
-31.6 &  90.06 & 0.40 & 0.38 & 3.08\\
-33.0 &  2.42 & 0.04 & 0.60 & 0.40\\
\hline
\multicolumn{5}{l}{G90.925+1.486 (MJD\textsubscript{s}=57885, $T\textsubscript{s}$= 5.379\,y, $N$=305)} \\
-69.2 &  61.49 & 0.36 & 0.44 & 3.99\\
-70.4 &  29.79 & 0.37 & 0.39 & 1.60\\
\hline
\multicolumn{5}{l}{G94.602-1.796 (MJD\textsubscript{s}=58027, $T\textsubscript{s}$= 4.996\,y, $N$=351)} \\
-40.9 &  5.30 & 0.14 & 0.45 & 0.68\\
-43.0 &  2.88 & 0.73 & 0.26 & 0.43\\
-43.7 &  3.74 & 0.43 & 0.47 & 0.59\\
\hline
\multicolumn{5}{l}{G173.482+2.446 (MJD\textsubscript{s}=57821, $T\textsubscript{s}$= 5.572\,y, $N$=250)} \\
-11.8 &  6.78 & 0.18 & 0.40 & 0.49\\
-13.0 &  25.46 & 0.36 & 0.29 & 1.57\\
-13.8 &  18.91 & 0.15 & 0.43 & 1.02\\
-7.5 &  9.86 & 0.41 & 0.73 & 3.78\\
\hline
\multicolumn{5}{l}{G111.26-0.77 (MJD\textsubscript{s}=57828, $T\textsubscript{s}$= 5.535\,y, $N$=315)} \\
-36.0 &  0.68 & 0.76 & 1.36 & 0.35\\
-38.0 &  1.64 & 0.59 & 0.66 & 0.69\\
\hline
\multicolumn{5}{l}{G111.542+0.777 (MJD\textsubscript{s}=57832, $T\textsubscript{s}$= 5.541\,y, $N$=276)} \\
-48.2 &  10.62 & 0.34 & 0.39 & 1.03\\
-48.8 &  9.82 & 0.40 & 0.40 & 1.15\\
-56.1 &  74.46 & 0.37 & 0.26 & 3.11\\
-56.8 &  103.97 & 0.14 & 0.31 & 0.62\\
-57.6 &  177.57 & 0.16 & 0.35 & 1.14\\
-58.0 &  247.68 & 0.19 & 0.31 & 0.72\\
-60.7 &  130.11 & 0.11 & 0.29 & 0.56\\
-61.3 &  165.87 & 0.20 & 0.44 & 1.70\\
\hline
\multicolumn{5}{l}{G133.947+1.064 (MJD\textsubscript{s}=57841, $T\textsubscript{s}$= 5.520\,y, $N$=297)} \\
-41.8 &  40.17 & 0.41 & 0.49 & 2.13\\
-42.2 &  289.10 & 0.62 & 1.25 & 10.33\\
-42.6 &  1376.89 & 0.29 & 0.33 & 2.45\\
-43.0 &  2230.47 & 0.12 & 0.36 & 1.25\\
-43.5 &  2800.11 & 0.11 & 0.33 & 1.15\\
-44.6 &  3366.63 & 0.11 & 0.33 & 1.14\\
-45.1 &  2612.50 & 0.12 & 0.35 & 1.27\\
-45.5 &  2090.53 & 0.09 & 0.29 & 1.02\\
\hline
\multicolumn{5}{l}{G109.871+2.114 (MJD\textsubscript{s}=57829, $T\textsubscript{s}$= 5.684\,y, $N$=1504)} \\
-1.8 &  225.67 & 0.45 & 0.72 & 18.04\\
-2.4 &  1026.05 & 0.59 & 0.83 & 50.19\\
-3.7 &  265.84 & 0.41 & 0.43 & 5.41\\
-4.0 &  418.37 & 0.18 & 0.30 & 1.17\\
-4.7 &  23.86 & 0.85 & 0.89 & 15.04\\
\hline
\multicolumn{5}{l}{G121.298+0.659 (MJD\textsubscript{s}=57841, $T\textsubscript{s}$= 5.605\,y, $N$=289)} \\
-22.4 &  1.86 & 0.67 & 0.44 & 0.64\\
-24.4 &  8.08 & 0.44 & 0.51 & 1.36\\
-24.9 &  7.72 & 0.85 & 1.00 & 10.55\\
-25.8 &  5.09 & 0.87 & 1.74 & 6.74\\
-27.0 &  1.70 & 0.34 & 0.51 & 0.39\\
-27.5 &  1.43 & 0.64 & 0.43 & 0.57\\
\hline
\multicolumn{5}{l}{G107.298+5.639 (MJD\textsubscript{s}=58298, $T\textsubscript{s}$= 4.285\,y, $N$=1823)} \\
-11.0 &  1.53 & 0.61 & 0.81 & 1.10\\
-16.7 &  1.71 & 0.91 & 3.09 & 1.96\\
-7.4 &  27.04 & 0.99 & 4.81 & 307.60\\
-8.6 &  4.30 & 0.88 & 1.76 & 8.21\\
-9.2 &  7.46 & 0.98 & 6.76 & 23.56\\
\hline
\multicolumn{5}{l}{G108.184+5.519 (MJD\textsubscript{s}=57839, $T\textsubscript{s}$= 5.523\,y, $N$=282)} \\
-10.0 &  17.20 & 0.58 & 0.36 & 4.33\\
-10.8 &  33.57 & 0.40 & 0.18 & 3.65\\
-12.7 &  15.56 & 0.36 & 0.47 & 1.12\\
\hline
\end{xtabular}


\bsp	
\label{lastpage}
\end{document}